\documentclass{emulateapj}

\usepackage{amssymb}
\usepackage{graphicx}

\slugcomment{To appear in ApJ, September 2010}

\begin{document}

\title{The Energetics of Molecular Gas in NGC 891 from H$_2$  and FIR Spectroscopy}

\author{G.J. Stacey\altaffilmark{1}, 
V. Charmandaris\altaffilmark{2,3},
F. Boulanger\altaffilmark{4},
Yanling Wu\altaffilmark{5},
F. Combes\altaffilmark{6},
S.J.U. Higdon\altaffilmark{7},
J.D.T. Smith\altaffilmark{8},
T. Nikola\altaffilmark{1}
}

\altaffiltext{1}{Astronomy Department, Cornell University, Ithaca, NY 14853, USA}

\altaffiltext{2}{University of Crete, Department of Physics, and Institute of Theoretical \& Computational Physics, GR-71003, Heraklion, Greece }

\altaffiltext{3}{IESL/Foundation for Research and Technology - Hellas,  GR-71110, Heraklion, Greece and Chercheur Associ\'e, Observatoire de  Paris, F-75014, Paris, France}

\altaffiltext{4}{Institut d'Astrophysique Spatiale, Universit\'e Paris Sud, B\^at. 121, F-91405, Orsay Cedex, France}

\altaffiltext{5}{Spitzer Science Center, MS 220-6, Caltech, Pasadena,  CA 91125, USA}

\altaffiltext{6}{Observatoire de Paris, LERMA, 61 Ave. de l'Observatoire, F-75014, Paris, France}

\altaffiltext{7}{Department of Physics, Georgia Southern University, Statesboro, GA 30460, USA}

\altaffiltext{8}{University of Toledo, Department of Physics \& Astronomy, Toledo, OH 43606, USA}

\email{stacey@astro.cornell.edu}

\begin{abstract}

We have studied the molecular hydrogen energetics of the edge-on spiral galaxy NGC\,891, using a 34-position map in the lowest three pure rotational H$_2$ lines observed with the Spitzer Infrared Spectrograph.  The S(0), S(1), and S(2) lines are bright with an extinction corrected total luminosity of $\sim2.8 \times 10^{7}$ L$_{\odot}$, or 0.09\% of the total-infrared luminosity of NGC\,891.  The H$_2$ line ratios are nearly constant along the plane of the galaxy -- we do not observe the previously reported strong drop-off in the S(1)/S(0) line intensity ratio in the outer regions of the galaxy, so we find no evidence for the very massive cold CO-free molecular clouds invoked to explain the past observations.  The H$_2$ level  excitation temperatures increase monotonically indicating more than one component to the emitting gas. More than 99\% of the mass is in the lowest excitation (T$_{ex}$ $\sim$125 K) ``warm'' component.  In the inner galaxy, the warm H$_2$ emitting gas is $\sim$15\% of the CO(1-0)-traced cool molecular gas, while in the outer regions the fraction is twice as high.  This large mass of warm gas is heated by a combination of the far-UV photons from stars in photo-dissociation regions (PDRs) and the dissipation of turbulent kinetic energy. Including the observed far-infrared [OI] and [CII] fine-structure line emission and far-infrared continuum emission in a self-consistent manner to constrain the PDR models, we find essentially all of the S(0) and most (70\%) of the S(1) line arises from low excitation PDRs, while most (80\%) of the S(2) and the remainder of the S(1) line emission arises from low velocity microturbulent dissipation.

\end{abstract}



\section{Introduction}

Half the interstellar medium in late type spiral galaxies is in
molecular gas clouds, and about 90\% by number of the atoms within
these clouds are tied up in the H$_2$ molecule. Stars form within
molecular clouds, and the pure rotational lines of H$_2$ are important
coolants enabling cloud collapse.  The cooling radiation from these
lines is also important to the formation of the giant molecular clouds
(GMCs) themselves. Despite its abundance and importance, very little
is known about cool H$_2$ on galactic scales.  This is because H$_2$
has no dipole moment, so that dipole radiation from its low-lying
rotational energy levels is forbidden.  The low lying levels instead
radiatively decay by emitting relatively weak quadrupole ($\Delta$J =
2) radiation. Therefore the pure rotational lines of H$_2$ are
challenging to observe.  Three of the low lying lines: the S(1)
(17.0$\mu$m), S(2) (12.3$\mu$m), and the S(4) (8.0$\mu$m) lines are
transmitted through the Earth's atmosphere, and have been detected
from regions as diverse as supernovae remnants \citep{Richter95},
photo-dissociation regions (PDRs) associated with OB star formation
sites \citep[e.g.][]{Parmar91}, and disks enveloping young stellar
objects \citep[e.g.][]{Bitner07}.  However, detection of the lowest
lying S(0) (28.2$\mu$m) line, and detection of these lines from cool
molecular clouds in external galaxies awaited the advent of the Short
Wavelength Spectrometers (SWS) on the Infrared Space Observatory
\citep[ISO, e.g.][]{Valentijn99}. A few years later, the Infrared
Spectrograph (IRS\footnote{The IRS was a collaborative venture between
  Cornell University and Ball Aerospace Corporation funded by NASA
  through the Jet Propulsion Laboratory and the Ames Research Center.
  Support for this work was provided by NASA through Contract Number
  1257184 issued by JPL/Caltech.}) of the Spitzer Space Telescope
became available and the detection of these lines from a wide variety
of external galaxies became routine
\citep[e.g.][]{Armus04,Smith04,Bernard09}.

H$_2$ exists in two varieties: para (nuclear spins anti-aligned) and
ortho (spins aligned).  The relative ortho-to-para ($o/p$) abundances
reflect the local gas temperature since proton exchange reactions with
H$^{+}$, H and H$_{3}^{+}$ are effective in thermalizing the $o/p$     ratio
\citep[][ and references therein]{Sternberg99}. At high temperatures
(T$>$200 K) the o/p ratio is set by the ratio of statistical weights
at 3:1. Since there is no radiative coupling between the ortho and
para species, when analyzing radiative transitions they can be treated
as independent molecules. The lowest lying transitions for the para
species are: J=2-0 (S(0), 28.2$\mu$m), and 4-2 (S(2), 12.3$\mu$m), and
for the ortho species: J=3-1 (S(1), 17.0$\mu$m), and J=5-3 (S(3),
9.7$\mu$m). With their very small Einstein A coefficients the H$_2$
level populations are easily thermalized at molecular cloud densities,
and the emitted lines are nearly always optically thin.  Therefore,
the line intensities trace the column density of the emitting level, and 
since the line emitting levels are thermalized and lie hundreds of degrees 
above the ground state (larger than typical molecular gas temperatures), the line ratios within a
species should reflect the gas temperature, T$_{gas}$. For example, the S(2)/S(0) line ratio
gets larger by a factor of 150 as T$_{gas}$ goes from 70 to 100 K.  The para J=2 level
lies just 510 K above the ground so it can be emitted by cool
(T$_{gas}>70$ K) molecular gas and is an important coolant for
molecular clouds in, for example, the early stages of collapse into
protostars (T$\sim$100 K, n$_{\rm H_2}\sim$100cm$^{-3}$).  

Are the rotational lines of H$_2$ detectable from normal star forming galaxies, and if so, is the emitting gas important?  Work with the ISO-SWS demonstrated that the answer to both questions is an emphatic yes.  Large scale S(0) and S(1) line emission is reported both from the face-on Sc galaxy NGC 6946 \citep[][]{Valentijn96}, and the edge on Sb galaxy NGC\,891 \citep[][]{Valentijn99}.  The strong emission indicates surprisingly large masses of warm molecular gas. Spitzer IRS studies confirm that strong H$_2$ rotational line emission and large masses of warm molecular gas are common in a wide variety of galaxies including nuclei of starburst galaxies \citep{Devost04, Bernard09} and ULIRGs \citep{Higdon06b}, in disk of galaxies \citep{Roussel07}, extragalactic regions such as Stephan's Quintet \citep{Appleton06}, as well as tidal dwarf galaxies \citep{Higdon06a}.

\object[NGC891]{NGC\,891} is a nearby (d = 9.5 Mpc \citet{vanderKruit}) Sb galaxy that is presented to us nearly edge-on and is thought to be a close Milky Way analogue.  Since an edge-on galaxy presents the largest column densities to our telescopes, and the H$_2$ rotational lines are optically thin, NGC\,891 provides the best opportunity to study warm H$_2$ in the outer reaches of galaxies.  With the ISO-SWS, \citet{Valentijn99} detected the S(0) and S(1) lines at eight positions along the plane of NGC\,891. The lines were surprisingly strong even in the outer regions of the stellar disk. Assuming $o/p$ = 3 they find warm (T$_{gas} \sim$130 to 230 K) molecular cloud component pervades the galactic disk with a total mass comparable to both the atomic HI and the cold molecular gas as traced through its CO emission: M$_{warm~H_{2}}$ $\sim $M$_{cold~CO}$ $\sim$ M$_{HI}$.  It would be a surprise if half the molecular ISM in the galaxy is warm, and it is a challenge to provide a heating mechanism for this gas.  Furthermore, based on a drop in the S(1)/S(0) line intensity ratio at radii greater than 8 kpc, \citet{Valentijn99} argue for an additional cool (T$_{gas} <$ 90 K) component which becomes dominant at large radii. With $o/p$ $\sim$2 to 3, the cooler component is massive: M$_{cool~H_{2}}$ $\sim$ 5-15 $\times$ M$_{HI}$ in the outer disk.  The cool molecular gas mass is so large that it can be a substantial fraction of the "dark matter" required to drive the rotation curve out to the edge of the opitical disk of NGC\,891. This is particularly intriguing due to theoretical work that suggests a large fraction of the ``dark'' baryonic matter in galaxies may be in CO free H$_2$ clouds \citep[e.g.][]{Pfenniger94}.  The detection of a massive cooler component, if proven correct, is a very significant result.  These results need verification so we have repeated the observations of the S(0) and S(1) lines along the plane of NGC\,891 using the more sensitive spectrograph on the Spitzer Space Telescope \citep{Houck04}.  In addition, the important S(2) line is included in the large instantaneous bandwidth of the Spitzer IRS.  The addition of the S(2) line mapping provides a key element in tracing the gas excitation.

\section{Observations}

The galaxy was observed on 2004 August 7 as part of the IRS
guaranteed time (GT) program \dataset[ADS/Sa.Spitzer\#0004935936]{(PID 97)} with the two high resolution (R$\sim$600) modules, Short-High (SH) and Long-High (LH). We also obtained supplemental, deep integrations in selected positions through a Spitzer Cycle 4 open time (OT) program with PID 40877. For all observations, the SH slits covers the 9.7-19.5$\mu$m range and it is 4.7$\arcsec\times11.3\arcsec$ in size, while LH covers the 18.8-37.2$\mu$m range and its size is 11.1$\arcsec\times22.3\arcsec$.  The width of each slit was designed so that it is equal to the first Airy ring of an unresolved point source at their short wavelength \citep[see][]{Houck04}. We obtained spectra for a total of 27 positions symmetrically sampled around the nucleus of the galaxy at a uniform spacing of 30$\arcsec$ (1.53 kpc) with a position angle of 23 degree along the galactic plane of NGC\,891. Six positions off the galactic plane to the northwest were also included to search for possible extended H$_{2}$ emission. For each pointing, the on-source integration time was 60 and 28 secs for SH and LH respectively. In addition two off positions to the northeast and northwest away from the galaxy, which were used for background subtraction, were also obtained with twice the integration time.  The IRS pointings are indicated in Figure 1 with circles of 25$''$ in diameter, slightly larger than the length of the LH slit. The enumeration of the pointings follows the notation of Table 1, which provides the RA and Dec of each pointing along with its distance in arcsec along the disk of the galaxy from the central pointing (ID=1).  Within the OT program, we obtained on-source integration times of 750 and 900 seconds for 9 positions (positions 1,6,8,10,12,20,22,24 and 26 in the notation of Figure 1) and 5 positions (positions 1,6,8,20,22) in SH and LH respectively.  We also re-observed the off possitions for background subtraction. 

Given the fact that IRS has no moving parts and that the long axis of the SH slit is perpendicular to that of the LH slit, we could not modify at will the orientation of the IRS slits on the sky. The slit orientation is determined by the ecliptic coordinates of the science target and the time of observations. By properly selecting the time of the observations, we were able to ensure that for the SH observations, which contain the S(1) and S(2) lines, the SH slit was placed with the long axis perpendicular to the plane of the galaxy. The LH observations, which trace the S(0) line, were obtained at the same time, with the LH slit placed with the long axis along the plane of the galaxy, since the two slits are perpendicular.  

The data were processed by the Spitzer Science Center (SSC) pipeline version 17.2. As it is discussed in detail in the IRS manual, for each pointing, two observations are made by placing the target on two nod positions along the slit. The 128$\times$128 pixel detectors are read with $N-1$ successive non destructive reads before the final n-th read which also resets the array. For our observations N=16 for SH and 8 for LH. The SSC pipeline starts with the 128$\times128\times N$ SH data cubes and converts them to a 128$\times128$ two-dimensional image after linearization correction, subtraction of darks, cosmic-ray removal, stray light and flat field correction. Each image, called basic calibrated data (BCD) contains the ten echelle orders of the SH and LH slits respectively. Our analysis started from these BCD files.  At each pointing, we subtracted the dedicated off-source background image in order to remove the sky emission from our slit. Then, the two dimensional images were cleaned with the IRSCLEAN package to remove the bad pixels as well as to correct the so called ``rogue" pixels. These are pixels whose values depend not only on the photons they collect, but also on the total number of photons hitting the whole 128$\times128$ array, in a nonlinear manner 
\footnote{For more details see the IRS manual at http://ssc.spitzer.caltech.edu/irs and http://ssc.spitzer.caltech.edu/dataanalysistools/tools/irsclean/} Finally, the cleaned images were used to extract the spectra using the full slit extraction method of the IRS data analysis package SMART \citep{Higdon04}. The outputs from SMART produced one spectrum at each nod position, and thene the two nod position spectra are combined. The data from the edge pixels within each echelle order were manually cleaned by truncating a few pixels at the beginning and end of each order and when there was inconsistency, the blue end of the order was preferred. No scaling was needed between the adjacent grating orders within the same module.  The final errors include both statistical and systematic effects, and we  report them in Table 1. The overall absolute spectrophotometric uncertainty of IRS is 5\%, so that spectrophotometric uncertainty is small compared with the signal-to-noise ratio of our spectra.

\section{Results}

\subsection{Line Morphology}

Table 1 lists the observed line fluxes and upper limits at all positions, while the locations on the galaxy where these spectra were obtained superposed on an 8$\mu$m Spitzer/IRAC image of the galaxy, are indicated in Figure 1. We also plot in Figure 2 the spectra around the S(0), S(1), and (S3) H$_{2}$ lines from locations where at least one line has been detected.  Both the S(0) and S(1) lines are clearly detected at 17 of the positions along the galactic plane. Figure 3a plots the line intensity distributions along the plane of galaxy. The lines have fairly uniform intensities in the inner galaxy, with some enhancement at the nucleus, beginning to fall off at 6.1 kpc NE and 4.6 kpc SW of the nucleus to values below our detection threshold at regions beyond 13.8 kpc NE and SW of the nucleus.  The S(2) line was only detected in 11 positions, all within 10 kpc of the nucleus, but this is very likely a sensitivity issue.  The detection limit is similar for all three lines, but the S(2) is typically 4 times fainter than the S(0) line and 1.7 times fainter than the S(1) line.  Therefore, its distribution could well be nearly identical to the lower J lines.  The line intensity ratios are also fairly constant across the galaxy, with the exception of enhanced S(1) line emission at the nucleus. 

We do not detect H$_2$ line emission off the plane of the galaxy. However, our integration time is rather modest, so perhaps this is not too surprising: \citet{Rand08} obtained substantially deeper integrations (nearly 6.5 hours, or 400 times longer than ours) at two positions 1 kpc off the disk, where they clearly detected the S(1) line.  Based on these results \citet{Rand08} estimate the scale height of the S(1) line emission is $\sim$8.4 to 9.9" (390-490 pc, based on a distance of 9.5 Mpc).  Correcting their results for the expected extinction of the lines by dust, they estimate that the intrinsic width of the warm molecular gas is $\sim$4.8 to 5.7$\arcsec$ (220 to 260 pc).     

\subsection{Comparison with ISO Results: No Evidence for a Massive  Cold Component}  

It is important to compare our Spitzer IRS results to those from the ISO SWS study reported in \citep{Valentijn96}.  Howevever both the ISO and Spitzer maps are sparsely sampled, and neither was obtained with a large enough beam to detect all of the extra-planar emission if it is as extensive as the results of \citet{Rand08} suggest.  We first correct for the missing extra-planar emission, by assuming the apparent (exponential) scale height for the H$_2$ line emission is the average of \citet{Rand08}'s values:  9.1$"$.  We took Spitzer LH (S(0) line) data with the slit width extending $\pm$5.55$"$ perpendicular to the plane, and SH (S1) and S(2) line data with the long axis of the slit expending $\pm$5.65$"$ perpendicular to the plane of the galaxy.  For these cases, we calculate that 45\% and 46\% of the line flux (integrated perpendicular to the plane) is detected within our LH and SH beams respectively.  The (rectangular) ISO beams were 20$" \times 27"$ and 14$" \times 20"$ for the S(0) and S(1) lines respectively, and these beams were tilted so that the long axis of the rectangle traced the plane of NGC 891 \citep{Valentijn96}.  For these cases, and again comparing against an assumed scale height of 9.1$"$ and integrated perpendicular to the plane, we calculate that 45\% and 53\% of the line flux was detected by the ISO SWS spectrometer.  We apply these extra-planar corrections to both data sets for proper comparisons.  We also must correct for flux not detected along the plane of the galaxy, since neither ours, nor the ISO data sets are fully sampled. The ISO SWS slit along the plane of the galaxy (point-to-point, along the diagonal) was 34$"$ and 24$"$ for the S(0) and S(1) lines respectively.  Since the Spitzer data are only sampled every 30$''$, we must interpolate between Spizer samplings to properly compare with the ISO data.  For both the S(0) and S(1) data sets, we interpolate linearly between the Spitzer sampled regions, then integrate within a region covering the angular extent of the ISO beam along the plane. This should be a good estimator for most of the galaxy, especially for the S(0) line, where the ISO and Spitzer beam sizes are nearly the same.  A possible exception is the nucleus in the S(1) line, where the relatively much smaller Spitzer beam may emphasize any strong central peak. Figure 3b illustrates the comparison between Spitzer and ISO observations including all the effects described above.  For the inner (r $<$ 7 kpc) galaxy our Spitzer line fluxes are about a factor 1.55 and 1.85 larger than those of ISO in the S(0) and S(1) lines respectively, but the Spitzer S(1)/S(0) line ratio is not much different than the ISO S(1)/S(0) ratio.  Therefore, we derive similar excitation temperatures for the inner galaxy as those presented in \citet{Valentijn96}(see section 4.1.1 below).  

However, in the outer galaxy (r $\geq$ 7 kpc) the S(1)/S(0) line ratios diverge.  The Spitzer S(1) line flux does not fall off nearly as rapidly as that measured by ISO, so that the Spitzer/ISO ratio grows to $\sim$ 2.57 in the outer regions.  In addition, the Spitzer S(0) line flux falls off more rapidly in the outer galaxy than does the ISO S(0) line flux, so that Spitzer/ISO ratio gets smaller, to $\sim$ 0.81 in the outer regions.  Therefore, unlike our Spitzer results, where the S(1)/S(0) line ratio is nearly constant over the disk, the ISO results show a marked decrease in the S(1)/S(0) ratio for the 3 regions sampled at r $\geq$ 7 kpc from the nucleus.  It is this change in line ratios for the outer regions that led \citet{Valentijn99} to invoke a second, cool gas contributor to the observed S(0) line emission.  Beam size is not the source of this discrepancy:  since the ISO beam has its diagonal along the plane of the galaxy, it was equally sensitive to extraplaner emission as the Spitzer beam.  Furthermore, it is not just the added S(0) line flux in the outer disk, but it is also the dearth of S(1) line radiation in the outer disk reported in the ISO observations that changes the observed S(1)/S(0) line ratio, hence the derived gas excitation.  The nearly constant line ratios revealed by the Spitzer observations are consistent with warm molecular gas, with near uniform excitation conditions over the galaxy - there is no evidence for an additional cool gas component.  Furthermore, as we will show below, this warm molecular gas component is far less massive ($<$13\%) than the cold H$_{2}$ emitting molecular component ($>~2.2 \times 10^9$ M$_{\odot}$) invoked by \citet{Valentijn99}.

\subsection{Extinction Correction}

To begin an analysis of mid-IR line emission from the edge-on galaxy NGC\,891, we need to first make corrections for the non-negligible extinction in the lines due to dust along the line of sight.  The H$_2$ line emission is likely to be highly correlated with that of the CO(1-0) line which is tightly confined to the plane of the galaxy, so that to make a good extinction correction, we need high spatial resolution proxies.  Fortunately, there exist high resolution images of NGC\,891 in the two primary tracers of the neutral ISM -- the HI 21cm line which traces atomic hydrogen, and the CO(1-0) line which traces cold molecular hydrogen.  We calculate the total column density in the molecular gas and atomic gas components within our Spitzer beams as a function of distance along the plane of the galaxy as follows.

For the CO(1-0) distribution, we use the data contained in Figure 2 of \citet{Scoville93}.  These data were obtained with a 2.3$''$ beam, but are plotted in their Figure 2 by integrating the emission perpendicular to the plane of the galaxy, then refering it to a 1$''$ strip along the plane.  To compare with our data, we therefore re-expand their strip to follow an exponential fall-off perpendicular to the plane of the galaxy with the local CO(1-0) scale height (also given in Figure 2 of \citet{Scoville93}). The observed scale height (given as full-width-at-half-maximum) varies and corresponds to a 1/e scale height ranging from about 2.9$''$ at the nucleus to $\sim$ 4.3$''$ in the outer regions.  For regions just exterior (further removed from the nucleus along the galactic plane) to those explored by \citet{Scoville93} (r$>$200$''$) we have added in the IRAM 30 m telescope (20$''$ beam) observations of \citet{Garcia92}, scaled assuming the line emission follows the 4.3$''$ scale height observed in the outer galaxy by \citet{Scoville93}.  We used a CO(1-0) integrated intensity to H$_2$ column density conversion factor of X = $2 \times 10^{20}$ cm$^{-2}$/(K km s$^{-1}$) as for the solar neighborhood \citep{Dame01}\footnote{A different choice of X will not change our results in any important way.  For example, if X = $3 \times 10^{20}$ cm$^{-2}$/(K km s$^{-1}$), then the extinction corrected luminosities of the H$_2$ lines go up 10\%, and the overall calculated warm (H$_2$ line emitting) to cold (CO(1-0) line emitting) molecular mass fraction goes down to 11\% from 16\% (see section 4.2.2). The derived physical parameters of the gas, and its excitation mechanisms do not change significantly.}, and derived an expected hydrogenic column density within 1$''$ strips parallel to the galactic plane with an exponential fall-off as given in \citet{Scoville93}.   

The HI column density is taken from the 21 cm line map of \citet{Rupen91} with a 7$''$ beam. Since the scale height of the HI emission is relatively large ($\sim$ 6$''$, \citet{Oosterloo07}, we estimate the HI column density within our Spizer beams by direct inspection of the HI maps. Note that since the HI column is a small fraction of the H$_2$ column (as traced by CO) in the the regions of the galaxy where extinction is important (within $\sim$ 5 kpc of the nucleus, see Figure 4a), a more detailed model of the HI distribution is not warranted.  


The total neutral gas mass is nearly equally divided into molecular ($\sim 3.8 \times 10^9$ M$_{\odot}$, Scoville et al. 1993), using X = $2 \times 10^{20}$ cm$^{-2}$/(K km s$^{-1}$) , and atomic \citep[$\sim 4.1 \times 10^9$ M$_{\odot}$][]{Oosterloo07} components. We plot the calculated column density of hydrogen nuclei derived from the CO and HI observations, smoothed to our S(0) beam size along the plane of the galaxy in Figure 4a.  The column plotted only includes mass within our Spitzer beam which extends $\pm$ 5.55$''$ perpendicular to the galactic plane:  the mass missed at high galactic latitude is not included in this plot. The total column density of hydrogen nuclei within our beam peaks at $1.28 \times 10^{23}$ cm$^{-2}$ at the nucleus and is above $2 \times 10^{22}$ cm$^{-2}$ from -10 to +10 kpc along the galactic plane.

Such large columns lead to sizable extinction even for the S(0) line at 28$\mu$m. To find the extinction, we take the column density of hydrogen nuclie as traced by the CO distribution (modeled as an exponential fall-off) and the observed HI distribution within our beam to calculate the expected extinction as a function of distance from the galactic plane.  The extinction calculation is broken up into 1$''$ strips parallel to the plane. For each strip, we first find A$_\lambda$, and then we calculate the local extinction correction, C($\lambda$) as defined below.  We then average these extinction corrections within our beam to get an estimate for the total correction that must be applied to the observed H$_{2}$ line emission.  Extinction is estimated using the ``Galactic'' dust model of \citet{Draine03} for which A(28$\mu$m)/N(H) $\sim 8 \times 10^{-24}$ mag\,cm$^2$, A(17$\mu$m)/N(H) $\sim$ A(12$\mu$m)/N(H) $\sim 1.43 \times 10^{-23}$ mag\,cm$^2$, and make the correction for the higher gas to dust mass ratio observed in NGC\,891 \citep[260, see][]{Alton00} as compared with the Galaxy \citep[160, see][]{Sodroski94}.  We assumed a ``mixed'' extinction scenario, where the dust is mixed with the warm H$_2$ gas, and for which the extinction correction, C($\lambda$) is given by C($\lambda$) = $\tau_{\lambda}$(1 - e$^{-\tau_{\lambda}}$)$^{-1}$.  At the nucleus we derive a (beam averaged) extinction correction for the S(0) line of C(28$\mu$m)$\sim$1.38, and for the S(1) and S(2) lines C(12$\mu$m), C(17$\mu$m)$\sim$1.71.  The extinction correction as a function of offset from the nucleus is plotted in Figure 4b.  The correction is significant for much of the inner (r < 5 Kpc) galaxy, and larger for the S(1) and S(2) lines, so that the extinction correction will change the line ratios.

\subsection{Line Luminosity}

The line fluxes change only gradually over the plane of the galaxy, and the slits likely enclose nearly all of the flux perpendicular to the plane.  Therefore, to find the total line luminosity, we simply integrate the line fluxes along the plane of the galaxy, interpolating in a linear manner between samplings.  For both the S(0) and S(1) lines, this amounts to integrating the observations and multiplying by the ratio of the sampling interval to the beam size along the plane: 30/22.3 = 1.35 and 30/4.7 = 6.38 for the S(0) and S(1) lines respectively.  The total line fluxes so obtained are F(S(0)) $\sim 1.95 \times 10^{-15}$ W m$^{-2}$, and F(S(1)) $\sim 3.20 \times 10^{-15}$ W m$^{-2}$.  At a distance of 9.5 Mpc, the corresponding line luminosities are $5.44 \times 10^6$, and $8.94 \times 10^6$ L$_{\odot}$ for the S(0) and S(1) lines respectively.  The S(2) line is not as well sampled as the S(1) line due to sensitivity limits, but assuming the S(2)/S(1) line ratio for the detected positions ($\sim$ 0.59) is the same throughout the galaxy, the S(2) line luminosity would be $\sim 5.27 \times 10^6$ L$_{\odot}$.  The total observed (and interpolated) power for these three low-lying lines is therefore $\sim 1.97 \times 10^7$ L$_{\odot}$, or $6.3 \times 10^{-4}$ of the total IR (TIR, 3-1100 $\mu$m) luminosity, $3.14 \times 10^{10}$ L$_{\odot}$ (flux densities from \citet{Brauher08}, TIR as defined in \citet{Dale02}).  The $\it{observed}$ low J molecular hydrogen lines are therefore $\sim$ 1.5 times as bright relative to the TIR continuum when compared to the sample of 57 normal galaxies surveyed by \citet{Roussel07} who find a typical ratio near $4 \times 10^{-4}$ with a scatter of about a factor of two.

Correction for extinction before integrating along the plane (factors of 1.16, 1.32, and 1.32 for the S(0), S(1) and S(2) lines respectively, section 3.3, above), which is bound to be more important for the edge-on NGC\,891 than for the typically less inclined galaxies in the \citet{Roussel07} sample, the total H$_2$ luminosity is $\sim 2.51 \times 10^{7}$ L$_{\odot}$ or $8.0 \times 10^{-4}$ of the TIR luminosity, or $\sim$ twice the typical values found in \citet{Roussel07}.

The question of line flux at high scale heights missed by the Spitzer beam is more difficult to quantify.  If the scale height derived by \citet{Rand08} applies the flux we observe is roughly equal to that we miss at scale heights greater than that of our beam ($\sim$ 5.6$''$).  The missed (high z) emission summed over the three lines is in this case therefore $\sim 1.97 \times 10^7$ L$_{\odot}$. The high-z emission would suffer very little extinction, so the the total luminosity in all three lines would be $\sim 2.51 + 1.97 = 4.48 \times 10^7$ L$_{\odot}$ in the case of a 9.1$''$ \citep{Rand08} scale height.  Another estimate is presented by presuming the H$_{2}$ scale height is the same as that of the CO(1-0) line.  Since the CO scale height is observed along the entire galactic plane, this is the correction we prefer.  For this case, the flux exterior to our beam (without extinction correction) is just 16$\%$ of the flux within our beam (before extinction correction) so that the total luminosity in all three lines is $\sim 1.97 \times 0.16 + 2.51 = 2.83 \times 10^7$ L$_{\odot}$.  Correcting for both extinction and extra-planar emission, the H$_2$ luminosity is therefore $\sim ~0.9$ to $1.43 \times 10^{-3}$ of the TIR luminosity. 

\section{Discussion}

\subsection{Intensity Ratios}

To derive the physical properties of the emitting gas, we need to compute ratios between lines measured within different apertures.  Inspecting Figure 3, it is clear that for the most part, the flux in the three lines changes gradually at the scale of LH slit length (22.3$''$) along the plane of the galaxy.
The S(1) and S(2) lines are observed with the long (11.3$''$) axis of the SH slit perpendicular to the plane so that it matches the z-sampling of the S(0) beam.  Therefore, to take ratios among the H$_{2}$ lines, we need only correct for the beam size along the galactic plane.  We do this by linearly interpolating between S(1) and S(2) samples (taken with a 4.7$''$ beam along the plane of the galaxy), and then estimating the average intensity expected in the 22.3$''$ S(0) beam from this interpolation. To obtain the physical parameters of the gas, we use the extinction corrected values of the fluxes contained within our Spitzer beams presented in Figure 5a. The extinction corrected line ratios are presented in Figure 5b. 

\subsection{Gas Excitation: Analytical Models}

\subsubsection{The Ortho to Para ratio}

The low lying rotational lines of H$_2$ are optically thin and easily thermalized so that the lines and their ratios probe the temperature and column density of the emitting gas. The S(2) and S(0) lines both arise from the para species of H$_2$ so that it in the high density limit (n$_{\rm H_2} > $n$_{crit}$(J=4) $\sim 10^3$ cm$^{-3}$, \citet{LeBourlot99}), the I$_{S(2)}$/I$_{S(0)}$ ratio is given by the para excitation temperature T$_{ex}$:

\begin{equation}
\frac{I_{\rm S(2)}}{I_{\rm S(0)}} = \frac{A_{\rm S(2)}}{A_{\rm S(0)}}
\frac{\nu_{\rm  S(2)}}{\nu_{\rm S(0)}}
\frac{g_{\rm S(2)}}{g_{\rm S(0)}}e^{-\Delta E/T_{ex}}
\end{equation}

where A$_{i}$ is the Einstein A coefficient for spontaneous emission of level i \citep[A$_{\rm S(2)}$ = 2.75$\times 10^{-9}$, A$_{\rm S(0)}$ = 2.94$\times 10^{-11}$, A$_{\rm S(1)}$ = 4.76$\times
10^{-9}$][]{Wolniewicz98}, $\nu_{i}$ is the frequency of transition, g$_i$ is the statistical weight of the level (g$_{\rm S(2)}$=9, g$_{\rm S(0)}$=5), and $\Delta$E is the energy difference between the
emitting levels (= 1172 K).  We plot the J=4 level excitation temperature so derived in Figure 6 (marked as S(2)/S(0)).  The excitation temperature is remarkably uniform over the disk at T$_{ex}$(2,0)=204$\pm$8.5\,K.

Between the para and ortho species, the line intensity ratios have the additional factor of the relative abundance, the ortho to para ratio, $o/p$.  For a high temperature (T $>$ 200 K) gas in local thermodynamic equilibrium, this ratio is given by the ratio of the statistical weights, of the two species, $o/p$ = 3.  Assuming the $o/p$ ratio is 3, then averaged over the eleven positions for which the S(2) line is detected (hereafter, the ``S(2) region"), the average extinction corrected S(1)/S(0) and S(2)/S(1) line intensity ratios (1.92 and 0.59) yield level excitation temperatures of T$_{ex}$(1,0) = 125 K, T$_{ex}$(2,0) = 204 K, and T$_{ex}$(2,1) = 388 K respectively.  The level excitation temperatures are monotonically increasing with J, which is permitted if there is more than one emitting components within the beam, each with a different gas excitation temperature.  For temperatures less than $\sim 200~$K, the equilibrium $o/p$ ratio is less than 3.  \citet{Burton92} calculate the equilibrium $o/p$ ratio as a function of gas temperature.  Making use of this work, we arrive at a self consistent solution for T$_{ex}$(1,0) = 134 K with $o/p$ = 2.30.  Since the S(1)/S(0) line intensity ratios are fairly uniform across the galaxy, we assume this $o/p$ ratio applies everywhere, and use it to compute T$_{ex}$(1,0).  Notice that T$_{ex}$(2,0) is not affected by the $o/p$ ratio, and T$_{ex}$(2,1) is much greater than 200 K, so that $o/p \sim 3$.  Figure 6 shows the computed excitation temperatures across the disk of the galaxy.  As we found for T$_{ex}$(2,0), T$_{ex}$(1,0) is also remarkably uniform across the plane with a mean value of T$_{ex}$(1,0)=132$\pm$6\,K. T$_{ex}$(2,1) shows more variation, but this is largely not statistically significant.  Excluding the data point 6 kpc to the NE of the nucleus, the mean value is T$_{ex}$(2,1) = 372 $\pm$41\,K.  The position at 6 kpc NE of the nucleus is statistically higher excitation, with T$_{ex}$(2,1) = 556 +127/-87 K.  

\subsubsection{Warm H$_2$ Gas Mass Fraction: Enhancement of the Northern Ring}

Since the line emission is optically thin, the line intensity yields the gas column of the emitting level, N$_{u}$: I = h$\nu _{\rm S(n)}$ A$_{\rm S(n)}$ N$_{u}/(4\pi)$, where A$_{\rm S(n)}$ is the Einstein A coefficient for the S(n) transition at frequency $\nu _{\rm S(n)}$ for n = 0,1,2.  Averaged over the eleven positions from which the S(2) line is detected, the column densities are N$_{2} = 1.22 \times 10^{20}$cm$^{-2}$, N$_{3} = 9.02 \times 10^{18}$ cm$^{-2}$, and N$_{4} = 6.81 \times 10^{17}$ cm$^{-2}$.  The total column density of molecular hydrogen is given by: N$_{tot}$ = (N$_{u}$/g$_{u}$) Z(T) e$^{-(E_{u}/(kT))}$ where g$_{u}$ = (o/p)/(2J+1), and Z(T) is the partition function:
\begin{equation}
Z(T) = (\frac{o}{p}) \sum_{odd-J}^\infty g_{J} e^{- \Delta E_{J}/T} + 
\sum_{even-J}^\infty g_{J} e^{- \Delta E_{J}/T} 
\end{equation}
The symbols have their meanings as before.  Since the level excitations are not all the same, there must be more than one warm molecular gas component.  The simplest model has two gas components, a ``warm" component, which will emit most of the observed S(0) line, and a ``hot" component that will emit most of the observed S(2) line. Both components will emit in the S(1) line. With a two component model there can be no unique solution as there are 4 variables (2 components, each with a T$_{ex}$ and N(H$_{2}$)), and just 3 observables (the lines).  We model the line emission averaged within the ``S(2) region".  The minimum mass solution will have a high gas temperature for the warm component.  The largest gas temperature permitted by the data is the observed 1-0 excitation temperature.  For this temperature, all of the observed S(0) and S(1) line emission arises from the warm component.  Therefore, there can be no self-consistent solution for all three H$_{2}$ lines since the warm component is too cool to emit much in the S(2) line, and hot component, which will emit most of the the S(2) line, will also emit in the S(1) line.  Using the partition function, and the relationship between line intensity and column density (I$_{J,J-1}$ = h$\nu$$_{J,J-1}$A$_{J,J-1}$N$_J$/(4$\pi$)), we have explored the gas temperature, column density parameter space for self consistent solutions, and find the minimum mass solution has gas temperatures of T$_{warm}$ = 127 K (o/p ratio shifts to 2.2), and T$_{hot}$ = 1700 K, and H$_{2}$ column densities of N(H$_{2,warm}$) = 3.8 $\times 10^{21}$ cm$^{-2}$, and N(H$_{2,hot}$) = 8.2 $\times 10^{18}$ cm$^{-2}$.  Within this model, 75$\%$ and 25$\%$ of the observed S(1) line emission arises from the warm and hot components respectively.  The warm component emits less than 5$\%$ of the observed S(2) line, while the hot component emits less than 1$\%$ of the S(0) line.  The case where 25$\%$ and 75$\%$ of the observed S(1) line emission arises from the warm and hot components respectively is obtained for gas temperatures of T$_{warm}$ = 107 K (o/p ratio shifts to 1.65), and T$_{hot}$ = 465 K, and H$_{2}$ column densities of N(H$_{2,warm}$) = 5.8 $\times 10^{21}$ cm$^{-2}$, and N(H$_{2,hot}$) = 3.4 $\times 10^{19}$ cm$^{-2}$.  For this case, less than 1$\%$ of the observed S(2) (S(0) line) arise from the warm (hot) component. For all cases, the warm component completely dominates (more than $99\%$) the H$_{2}$ column density.  

We use the minimum mass self consistent model (75/25 split of S(1) line flux into warm and hot components) to calculate the total column density in H$_{2}$ line emitting molecular gas (esssentially the warm component from above) across the disk of NGC\,891.  The results, along with the column traced in CO(1-0) are plotted in Figure 7a.  The warm molecular gas column peaks at the nucleus at a value of $\sim 5.26 \times 10^{21}$ cm$^{-2}$, and is above $2 \times 10^{21}$ cm$^{-2}$ out to $\pm$ 8 kpc from the nucleus.  Figure 7b shows the ratio of the warm molecular gas to that traced in its CO(1-0) line emission (within the Spitzer beam), i.e. the fraction of the total molecular gas that is emitting in the H$_{2}$ lines.  This fraction is modest ($\sim$13\%) and nearly constant in the inner galaxy from 3 kpc NE of the nucleus to 6 kpc SE of the nucleus, but rises to $\sim$ 30\% in the outer galaxy.  This could be a true enhancement of the warm molecular gas mass in the outer galaxy, or it could be due to a change in the conversion factor, X between CO intensity and H$_2$ column density with distance from the nucleus.  If the X factor were larger by a factor of two in the outer galaxy compared with the inner galaxy, at first glance, this could explain our apparent change in the warm/cold molecular gas fraction.  However, we do not believe this is the case, as there is a significant NE-SW asymmetry in the warm/cold molecular gas fraction apparent within the molecular ring (r = 3 to 7 kpc).  Averaged over the ring positions, the NE portion of the ring has twice the warm molecular gas fraction ($\sim$ 28\%) as the SW portion of the ring ($\sim$14\%).  The total molecular gas as traced by CO is very close to the same in both sides of the ring \citep{Scoville93}: it is the warm molecular gas column that is enhanced (Figure 7, top).  To cancel out this effect, one would need to invoke an X factor twice as large for the NE ring as for the SW ring. This enhanced warm molecular mass fraction is more likely the result of enhanced starformation in the NE ring, which is reflected in the stronger [CII] and [OI] line emission observed to the NE as well (see Figure 8).  Integrated over the galaxy, the warm molecular gas totals $\sim6.1 \times 10^{8}$ M$_{\odot}$, or about 16\% of the total molecular gas mass as traced by CO ($3.9 \times 10^{9}$ M$_{\odot}$) within our Spitzer beams (which extend $\pm$5.5$''$ of the galactic plane). 

\subsection{Gas Heating}

The power to heat large amounts of molecular gas to temperatures above a hundred degrees can be provided by far-UV (stellar) photons, by the dissipation of molecular gas turbulent kinetic energy in cloud-cloud collisions and on smaller scales within clouds, by X-ray  photons, or by cosmic rays.  For normal star-forming galaxies, only the first two may provide enough power to be important.

\subsubsection{Photodissociation Regions -- Motivation}

Figure 8 provides insights into the source for the heating of the molecular gas. In this figure we plot the ``cold" molecular gas mass (traced in its CO emission) the ``warm" molecular gas mass (traced in its H$_2$ line emission), and the [CII] 158 $\mu$m and the [OI] 63 fine-structure lines obtained with the ISO LWS spectrometer \citep{Brauher08}.  We also include weak detections of the [OI] 146 $\mu$m at 6 positions along the galactic plane that we obtained through close inspection of the LWS L02 scans of the line obtained within the ISO data archive.  These data were calibrated as described in \citet{Brauher08} so as to be consistent with their [CII] and [OI] 63 $\mu$m fluxes, and their [NII] 122 $\mu$ fluxes which we use below.  The cold and warm molecular gas distributions have been smoothed to the significantly coarser spatial resolution ($\sim$75$''$) of the ISO LWS beam for proper comparisons, and the tracers are normalized to their nuclear values (excepting the [OI] lines which are normalized to their peaks) for ease of comparision. The cold molecular gas mass is centered on the nucleus with a significantly narrow distribution along the galactic plane than the warm molecular gas mass.  In addition, the warm molecular gas mass distribution is asymetrical, with near nuclear values extending to the molecular ring, 4.6 kpc to the NE of the nucleus. 

The [CII] line predominantly arises from photodissociation regions formed on the surfaces of molecular clouds exposed to the far-UV (6\,eV $< h\nu < 13.6\,$eV) radiation from nearby OB stars or the general interstellar radiation field.  The [CII] line is also an important coolant for low density ionized gas, and ``atomic clouds", but most \citep[$>$ 70\%][]{Stacey91, Abel05, Oberst06} of the [CII] flux from external galaxies likely arises from PDRs, so that the line is an indicator of OB star formation activity \citep{Stacey91}.  The very close correspondance between the [CII] distribution and that of the warm molecular gas is strongly suggestive of a PDR origin for the H$_{2}$ line emission.

The ionization potential for O (13.62 eV) is essentially the same as that of H, so that the FIR [OI] 63 and 146 $\mu$m line emission also arises from neutral gas within PDRs.  The [OI] lines have significantly higher critical densities ($\sim$ 4.7 $\times$ 10$^{5}$ and 9.4 $\times$ 10$^{4}$ cm$^{-3}$ for the 63 and 146 $\mu$m lines respectively) than that of the [CII] line ($\sim$ 2.8 $\times$ 10$^{3}$ cm$^{-3}$) so that the [OI] lines trace denser PDRs. The strong peaking of the [OI]63 $\mu$m line line on the northern molecular ring is strongly suggestive of enhanced gas densities in this region. 

The remarkable spatial correlation between the warm H$_2$ mass and the [CII] line including the extension to the NE suggests a common origin in PDRs.  Indeed, warm molecular gas is expected within PDRs.  The penetration of carbon ionizing photons into neutral gas clouds is limited by the extinction of these photons by dust to visual extinctions, A$_{\rm V} \sim 3$ mag (corresponding to an N$_{\rm H} \sim 6 \times 10^{21}$ cm$^{-2}$), and along half this column, hydrogen is typically molecular in form \citep{Tielens85}.

\subsubsection{Photodissociation Regions: Constraints on G$_{0}$}

We have argued for a PDR orgin for much of the H$_{2}$ line emission.  As such, one might expect a correlation between the observed H$_2$ line emission and that of an independent tracer of PDRs, the PAH 7.7 $\mu$m emission feature (e.g. \citet{Habart04},\citet{Peeters04}.  In Figure 9, we compare the PAH 7.7$\mu$m emission with the H$_2$ line emission, by plotting the H$_2$-to-PAH emission ratio along the plane of NGC\,891. The PAH emission is measured from IRAC 8 $\mu$m imaging data \citep{Whaley09} using the same apertures than those used for the H$_2$ spectroscopy. To compute the PAH emission from the 8 $\mu$m flux, we use the same effective bandwith ($\Delta \nu = 13.9\times 10^{12}\,{\rm Hz}$ ) as \citet{Roussel07}. The H$_2$ line emission is the sum of the S(0), S(1) and S(2) lines. All tracers were corrected for extinction (see section 3.3).  The extinction at 7.7 $\mu$m is similar to that at 12.3 and 17 $\mu$m \citep{Draine03}, so that the H$_2$ to PAH emission ratio is not very sensitive to the specific choice of the extinction curve. \citet{Rigopoulou02} and \citet{Roussel07} have shown that the H$_2$ emission of star forming galaxies is tightly correlated with the PAH emission. Figure 9 extends this correlation to the distribution of both tracers within a galaxy. The H$_2$ to PAH emission ratio is remarkably constant within the NGC\,891 disk, with a mean value only slightly larger than that measured for the SINGS galaxies.  

A striking outcome of the SINGS data analysis is that the H$_2$/PAH emission ratio is independent of the dust heating as measured by the $24~\mu$m to FIR dust emission ratio. \citet{Roussel07} argue that this result is easiest to understand if the H$_2$ and dust emission come both from PDRs, and are powered by a common energy source, the UV light from young stars.  However, from their discussion it is not fully clear how this interpretation fits with the modeling of the dust spectral energy distribution in \citet{Draine07}, where the PAH emission comes mainly from interstellar matter in a low UV radiation field (typically G $<10$). Here, we quantify the constraint set by the H$_2$/PAH emission ratio on the UV field.  Using the \citet{Draine_Li07} models available on line, we computed that the PAH emission, as it appears in Figure 9, is 

\begin{equation}
\rm{
L_{PAH} = 0.083 \, G \left( \frac{q_{PAH}}{3.6\%} \right) \left( \frac{M_H}{M_\odot} \right)  L_\odot
}
\end{equation}

where the dus-to-gas mass ratio is assumed to be Galactic and q$_{\rm PAH}$ is the fraction of the dust mass in PAHs \citep{Draine_Li07}. The default value of q$_{\rm PAH}$ corresponds to the median value derived by \citep{Draine07} for metallic galaxies in the SINGS sample. The model output may be combined with the observed value of H$_2$/PAH ($\sim$ 0.012), to estimate the effective radiation field, G,  within the warm H$_2$ gas. For NGC 891, we find L(H$_2$)/M(H$_2$,warm) $\sim$ 1/20, so that putting this all together we have: 

\begin{equation}
\rm{
G < 50 \left( \frac{3.6\%} {q_{PAH}} \right) \left( \frac{M_{H_{2}Warm}}{6.1\times 10^8\, M_ \odot} \right)^{-1}}
\end{equation}

Here we have left in an explicit scaling factor for the warm molecular gas mass.  Clearly, this calculation only provides a rough upper limit for G, since it assumes that the PAH emission arises exclusively from PDRs containing the  warm H$_2$ gas, while some PAH emission is expected from the H~I and cold H$_2$ gas. However, within the current understanding of PAH emission, the H$_2$ to PAH comparison therefore suggests that the bulk of the warm H$_2$ mass (the S(0) emitting gas) is located in PDRs exposed to modest far-UV radiation fields.  We use this as a constraint in our PDR models that we develop below.   It is important to note that this constraint on G only applies for the gas that radiates in the S(0) and S(1) lines, not the gas radiating in the S(2) line.  The S(2) line emission could arise from a much higher G environment without increasing the PAH flux very much since the mass of this component will be so small.

\subsubsection{Photodissociation Regions: Modeling}
\citet{Kaufman06} have made available a grid of PDR model outputs, which include the expected rotational line emission from H$_2$ as a function of G, the strength of the local far-UV (6 $< h \nu < 13.6$ eV) radiation field normalized to the Habing field (G$_{0} = 1.6 \times 10^{-3}$ erg s$^{-1}$ cm$^{-2}$), and the cloud number density n.  The model computes the local heating and cooling, chemistry and radiative transfer as a function of depth from the cloud surface into its dust-shielded core.  Diagnostic line ratios and line intensities appropriate for a single face-on slab geometry are posted on-line, and available interactively in the ``Photodissociation Region Toolbox (PDRT)\footnote{See dedicated web site at http://dustem.astro.umd.edu}".  Table 2 lists our observed H$_2$ S(0), S(1) and S(2) luminosities and those of the [CII], and [OI] lines integrated over the galaxy.  We estimate that the errors for the H$_2$ luminosities are roughly given by their corrections for extinction, and high latitude flux missed by the Spitzer beam.  Figure 10 (left) plots our observed galactic average H$_2$ line ratios as a function of G and n from PDRT.  The model results do not give a good fit to the NGC\,891 line ratios.  The best fit is for high density (n$\geq6 \times 10^4$ cm$^{-3}$) and moderate far-UV radiation fields (G $\sim$ 200 to 400).

Figure 10 (center) plots the [OI], [CII] and ([OI] 63~$\mu$m + [CII])/FIR continuum ratios as averaged over the galaxy onto a PDR diagnostic diagram from PDRT. For this FIR data set we make corrections to the observed luminosities of our tracers.  The first correction is for optical depth and geometry. The PDRT models are appropriate for face-on single slab geometry. Since the ISO beams encompass regions $\sim$ 2.5 kpc in size at the distance to NGC\,891, there is an ensemble of clouds within our beam.  For a spherical cloud, the projected surface area is only 1/4 the total surface area, so that a beam filled with spherical clouds will have 4 times the intensity as a unit filling factor single slab if the lines are optically thin.  If all the tracers are optically thin, taking the line ratios eliminates this geometric factor.  Although it is almost certainly true that all the H$_2$ lines are optically thin, optical depth may be an issue for the FIR lines.  In fact, the 63~$\mu$m [OI] is most likely optically thick \citep{Stacey83, Tielens85} in star formation regions, so that the observed [OI] line intensity should be corrected for geometry, multiplying by as much as a factor of four for high optical depth in the line in spherical geometry.  For our study, we make a more modest correction, multiplying the 63~$\mu$m line by a factor of two.  We take the [OI] error boundary to be 50$\%$ due to this factor of two correction, and note that this error bound will encompass geometry effects up to a factor of 3.  The [CII] and [OI] 146 $\mu$m lines are likely optically thin, or thinish as is the FIR continuum, so we make no optical depth correction to their fluxes.

Next, we correct the observed [CII] line emission for that fraction expected to arise from diffuse ionized gas.  This correction is perhaps best done by comparing the observed [CII] line flux to that of the ground state fine structure lines of N$^{+}$ at 205 and 122 $\mu$m.   With an ionization potential of 14.5 eV, these lines only arise from within ionized gas regions, and one can show that the [CII]/[NII] line ratios are a measure of the observed fraction of the [CII] line that arises from the ionized interstellar medium \citep[cf.][]{Petuchowski93, Heiles94, Oberst06}. We use the [NII] 122 $\mu$m lines observed by the ISO LWS \citep{Brauher08}, averaged over the plane of the galaxy.  The [NII] 122$\mu$m line to [CII] line ratio in NGC\,891 is $\sim$1:9: somewhat smaller than that of the Milky Way Galaxy \citep[1:6,][]{Wright91}. Assuming Galactic C/N abundance ratios the line ratio indicates that between 50\% (low density limit) and 4\% (high density limit) of the observed [CII] comes from ionized gas regions \citep{Oberst06}. Here we will assume 27\% of the [CII] line arises from HII regions as found for the low density (n$_e$$\sim$ 30 cm$^{-3}$) gas in the Carina nebula\citep{Oberst06}. We take the error bound for the [CII] luminosity to be equal to the fraction (27$\%$) we have subtracted off.

Finally, not all of the FIR continuum luminosity from galaxies will arise from dense PDRs.  A significant fraction can arise from the diffuse ISM.  \citet{Alton98} and \citet{Popescu04} have fit ``warm" (PDR) and ``cool" (diffuse) fractions of the total FIR luminosity from NGC\,891.  The warm dust likely arises from H~II regions and PDRs, while the cool dust arises from the cold interiors of molecular clouds and the diffuse ISM \citep{Draine07}. We therefore take $\sim$50$\%$ of the total FIR luminosity of NGC\,891 to arise from PDRs.  Finally, we correct L$_{FIR}$ to L$_{30 - 1000 \mu m}$ which is appropriate for the PDRT models (see footnote to Table 2). Considering these corrections, we take the error bound for the FIR luminosity to be 50$\%$.  Putting the FIR line and continum corrections together, the PDR cooling line/cooling continuum ratio integrated over NGC\,891 is ([OI] + [CII])/FIR $\sim$1.7\% (Table 2).

Let us now examine the overall properties of NGC\,891, integrated along the galactic plane.  Taken within themselves, the S(2)/S(0) and S(1)/S(0) line ratios (1.08 and 1.69 respectively) indicate G $\sim$200-400, and n $\geq$6 $\times$ 10$^4$ cm$^{-3}$. However, the [OI] 146$\mu$m/[CII], [OI] 63$\mu$m/[CII] and ([OI]+[CII])/FIR ratios indicate lower far-UV fields with best fit near G$\sim$100, and a significantly lower density with best fit near n$\sim$ 3$\times 10^3$ (Figure 9, center). The solution spaces for the H$_{2}$ and FIR line models do not overlap.  The closest approach for the two models is obtained for similar Gs ($\sim$ 100 to 200), but the FIR lines require a lower gas density (n$\sim$ 10$^{4}$ to 4 $\times 10^{3}$ cm$^{-3}$, as G ranges from 100 to 200) than the H$_2$ lines (n$> 6 \times 10^{4}$ cm$^{-3}$).  If we imagine the PDRs to be the surfaces of externally illuminated molecular clouds, then since the H$_2$ lines arise from deeper within the PDR than the FIR lines, it is likely they will arise from denser gas regions, so perhaps this is an acceptable solution.  If so, then the beam filling factors of the two components should match as well.  For the FIR component, the average [CII] line intensity (less 27\% for ionized gas) referred to an 11$''$ region along the plane of the galaxy is $2.8 \times 10^{-4}$ erg s$^{-1}$ cm$^{-2}$ sr$^{-1}$.  At the closest approach solution between the H$_{2}$ and FIR models (n$\sim 4\times 10^{3}$ cm$^{-3}$ and G$\sim$200), the [CII] intensity predicted by PDRT is $\sim 2.2 \times 10^{-4}$ erg s$^{-1}$ cm$^{-2}$ sr$^{-1}$ so that the beam filling factor for the FIR lines is $\phi_{\rm FIR~lines}$ $\sim$1.3.  For n$\sim 6 \times 10^{4}$, G$\sim$200, PDRT predicts a S(0) line intensity $\sim 1.0 \times 10^{-5}$ erg s$^{-1}$ cm$^{-2}$ sr$^{-1}$.  The extinction corrected value within our 11$''$ beam averaged over the plane of the galaxy is $\sim1.7 \times 10^{-5}$ erg s$^{-1}$ cm$^{-2}$ sr$^{-1}$, so that the beam filling factor is $\phi_{\rm H_2~lines}$ $\sim$ 1.7, in fair agreement with the FIR value.

\subsubsection{Is some of the H$_2$ emission powered by the dissipation of turbulence?}

The H$_{2}$ lines, the FIR lines, and the FIR continuum are consistent with PDR models but only if there is a rather large (factor of $\sim$10) density gradient between the (denser) H$_2$ emitting region and the (less dense) FIR line emitting regions.  In addition, the requisite far-UV field strength is significantly larger than the value we derive using the observed PAH distribution in section 4.3.2 above.  These difficulties motivate us to investigate the possibility that a fraction of the observed H$_2$ line emission arises from the dissipation of turbulence in molecular clouds. This possibility has been proposed to account for diffuse H$_2$ line emission in the Milky Way \citep{Falgarone05} and the galaxy-wide shock in Stephan's Quintet  \citep{Guillard09}.
 
The S(1)/S(0) line ratio is fully consistent with the far-UV fields and densities as derived from the FIR lines and continuum (see Figure 10, left). What drives the H$_2$ solution is the S(2) line, which is a factor of six too bright. The C-shock models of \citet{Draine83} and, more recently \citet{Flower10} are in good agreement, and indicate that the S(2) line is brighter than the S(1) line for v$_{\rm shock}$~$\geq$ 15 km s$^{-1}$.  For example, using the \citet{Flower10} model, for shock velocities, v$_{\rm shock}~\sim$20 to 30 km s$^{-1}$, with preshock density n$_{H}$ = n(H)+2n(H$_2$)= 2$\times$10$^4$ cm$^{-3}$, 
the S(2) line is $\sim$ 1.6 $\times$ as bright as the S(1) line, and 75 times brighter than the S(0) line.  If 80\% of the observed S(2) line flux were to arise from such C-shocks, and the remainder from PDRs, then 30\% of the S(1) line would also arise from shocks, and the PDR line ratios would shift to S(2)/S(1) $\sim$0.17, and S(2)/S(0)$\sim$ 0.22.  The allowed values for G and n for this PDR plus shocks solution are shaded blue in Figure 10 (right).  Clearly there is now a large region of overlap between the PDR solution for the FIR lines, and that for the H$_{2}$ lines.  This solution space is centered at about G$\sim$60 $\pm$ 40, n$\sim$ 5 +5/-3 $\times 10^{3}$ cm$^{-3}$ (Figure 10, center).  The revised filling factors are reasonably self-consistent with $\phi_{\rm FIR~lines}$ $\sim$ 1.9, and $\phi_{\rm H_2~lines}$ $\sim$ 2.8.  With v$_{\rm shock} \sim$ 25 km s$^{-1}$, there is significant cooling in other molecular hydrogen lines: the S(3) and S(4) lines are predicted to be 4$\sim$7 and 4 times brighter respectively than the S(2) line.  The total luminosity in the molecular hydrogen lines from the shocked gas would be very large: L$_{\rm H_2} \sim 10^8$ L$_{\odot}$.  For a 25 km s$^{-1}$ C-shock, the common cooling lines of [OI] and CO are more than 100 and 15 times less important energetically than those of H$_2$ , and H$_2$O is predicted to have a total cooling $\sim$ that of H$_2$ \citep{Flower10}. A similar analyis of the region centered on the north-east ring has a solution centered on modestly enhanced excitation, G$_{NE~ring} \sim$80, n$_{NE~ring} \sim 8 \times 10^{3}$ cm$^{-3}$, and requires somewhat higher area filling factors $\phi_{\rm FIR~lines}$ $\sim$ 2.7, and $\phi_{\rm H_2~lines}$ $\sim$ 3.6 to account for the enhanced [OI] 63 $\mu$m line emission there.

Leaving aside the question of the H$_2$ excitation, we can assess the contribution of mechanical energy dissipation to the H$_2$ emission, from the point of view of energetics.  Is the dissipation rate of the interstellar medium turbulent kinetic  large enough to provide a significant contribution to the H$_2$ emission?  Various groups have discussed the dissipation of turbulence using numerical magneto-hydrodynamic simulations \citep[e.g.]{Stone98,Mac99}. As in \citet{Bradford03}, one can rearrange equation (7) from \citet{Mac99} to parameterize the energy released by the dissipation of turbulence per unit mass in terms of the dispersion of the  turbulent velocity, v$_{rms}$, and the scale
length of energy injection, $\Lambda_{d}$: 

\begin{equation}
\frac{\rm L}{\rm M} = 2.5\times 10^{-3} \left( \frac{\rm v_{rms}}{10 \rm km s^{-1}} \right) ^{3}
				\left( \frac{\rm 100pc}{\Lambda_{d}} \right)  
				\frac{\rm L_{\odot}}{\rm M_{\odot}} 
\end{equation}

The default values for v$_{rms}$ and $\Lambda_{d}$ correspond to the upper end of the line width-size relation for giant molecular clouds in the Galaxy \citep{Solomon87}. We invoke shocks to account for 80\% of the S(2) luminosity, and 25\% of the S(1) line luminosity.  By dividing the sum of these line luminosiites $\rm 9.5\times 10^6 \, L_\odot$, by the molecular mass $\rm 4.4 \times 10^9 \, M_\odot$, derived from the CO(1-0) luminosity, we obtain $\rm L_{\rm{H_2}}/M_{\rm{H_2}} = 2.2 \times 10^{-3} \frac{\rm L_{\odot}}{\rm M_{\odot}} $.   Therefore, the expected turbulent dissipation rate is greater than the luminosites of the lines, so that a significant contribution from turbulent energy dissipation is likely.

In summary, the observed H$_{2}$ line emission from NGC 891 is well fit by a model that combines a low excitation PDR (G $\sim$ 60, n$\sim$5$ \times$ 10$^{3}$ cm$^{-3}$) with dissipation of turbulent energy within molecular clouds.  The (adjusted) [OI] (63 and 146 $\mu$m) and [CII] 158 $\mu$m fine structure line fluxes, (half) of the observed FIR continuum flux, the H$_{2}$ S(0) line flux and 70$\%$ of the S(1) line flux arise from the PDR component, while the H$_{2}$ S(2) line and 30$\%$ of the S(1) line flux arise from gas heated by turbulent dissipation.  It is interesting to note that based on observations of the H$_2$ lines along a line of sight within the Milky Way galaxy, \citet{Falgarone05} reached much the same conclusion:  the S(0) line emission may be accounted for by UV heated H$_2$ in the diffuse interstellar medium and low excitation PDRs, but the UV emission cannot be the unique energy source, because their PDR models fall short of the observed S(1) and S(2) lines by large factors ($\sim$2-10). As we outline here, they propose that the S(1) and S(2) H$_2$ line emission is powered by the localized dissipation of turbulent kinetic energy, albeit in a low density gas (n$_{\rm H}\sim 50\, {\rm cm^{-3}}$) within the diffuse interstellar medium.  The presence of a diffuse component of H$_2$ line emission has been confirmed by additional Galactic observations \citep[]{Hewitt09,Goldsmith10}. 

It should be clear that the solutions presented here are by no means unique.  On galactic scales there will certainly be a range of PDRs along the line of sight with varying degrees of excitation, that could mimic the observed line ratios, perhaps negating the need to invoke shocks.  Furthermore, a single parameter shock model is also simplistic: there will very likely be a superposition of shocks of various parameters only any given line of sight.  What we present here is intended to be only a representative solution.

\section{Conclusions}

We have mapped the distribution of the S(0), S(1), and S(2) pure rotational lines of H$_{2}$ along the plane of the edge-on galaxy NGC\,891. The lines are remarkably bright.  After making an extinction correction, the total line flux from the three transitions is $ 2.82\times 10^7$ L$_{\odot}$ or $\sim$0.09\% of the TIR luminosity of the galaxy.  Unlike the ISO results that suggest a change in the S(1)/S(0) line intensity ratio in the outer regions of the galaxy, we find the line ratios are nearly constant along the plane of the galaxy indicating uniform excitation of the gas.  In particular, there is no reason to invoke a massive cold component of the molecular gas in the outer galaxy as the ISO results appear to require.  Through analytical modeling, we find the bulk of the mass traced in its H$_2$ line emission has an excitation temperature near 125 K.  This ``warm'' molecular component is a substantial fraction of the total molecular gas mass: about 10 to 15\% in the inner (3 kpc NE $<$ r $<$ 6 kpc SW) galaxy rising to more than 30\% in the outer regions. Integrated over the galaxy, M$_{warm~H_{2}}$ $\sim$ 16\% M$_{cold~CO}$.

We compare our H$_2$ observations with detailed PDR models to show that the observed line intensities and ratios are consistent with high density (n$\geq 6 \times 10^4$ cm$^{-3}$) molecular clouds
exposed to moderate far-UV fields (G$\sim$200-400).  We refine our PDR modeling by using available ISO [OI] and [CII] data, which indicate smaller far-UV fields (G$\sim$5-300), and lower gas densities (n$\sim 2\times 10^{4} - 10^3$ cm$^{-3}$).  The solutions spaces for the H$_2$ and FIR line PDRs do not overlap, but a reasonably close solution is obtained for G$\sim$ 200, and a large (factor of ten) density gradient between the FIR emitting gas and the H$_2$ emitting gas.  This solution may be possible if the clouds are externally illuminated, but the high field invoked by this solution is in contradiction to the more modest far-UV fields (G$\sim50$) one would expect by comparing the observed H$_2$ line luminosity to that of the 7.7 $\mu$m PAH features. A consistent solution is obtained within a two component model.  The first component is a low excitation PDR, with G $\sim$ 60, and the second component is heated by the dissipation of turbulent kinetic energy.  We find a very good fit for the H$_2$ and FIR lines and continuum if all the observed S(0) line emission arises from the PDR component, and 80$\%$ of the S(2) line arises from the turbulent component.  The S(1) line emission is split 70\% into the PDR component, and 30\% into the turbulent component. This solution is only representative of a range of solutions that could be obtained by allowing more PDR and/or shock components with different excitation parameters.  

This is the first analysis of the H$_2$ rotational line emission from a galaxy together with the FIR lines and continuum in the context of PDR models.  A complication with this modeling is the relatively large beams and modest sensitivies of the ISO spectrometers for detection of the FIR lines.  These issues will become much less important in the near future with the arrival of FIR data from the 3 m class Stratospheric Observatory for Infrared Astronomy (SOFIA) and the Herschel Space Telescope spectrometers and photometers.     

\acknowledgments 

We thank an anonymous referee for many inciteful comments that led to a much improved version of this manuscript.  This work is based in part on observations made with the Spitzer Space Telescope, which is operated by the Jet Propulsion Laboratory, California Institute of Technology, under NASA contract 1407. Support for this work was provided by NASA through Contract Numbers 1257184 and 1311091 issued by JPL/Caltech. VC  acknowledges partial support from the EU ToK grant 39965 and FP7-REGPOT 206469.

\newpage

\begin{deluxetable}{rcccccc}
  \tabletypesize{\scriptsize}
  \setlength{\tabcolsep}{0.02in}
  \tablecaption{H$_2$ Line measurements along NGC\,891\label{tab1}}
  \tablewidth{0pc}
  \tablehead{
    \colhead{ID}   & \colhead{RA}    & \colhead{Dec}  & \colhead{Distance\tablenotemark{a}} & 
    \colhead{S(2) $\lambda$12.3$\mu$m} & \colhead{S(1) $\lambda$17.0$\mu$m}  & \colhead{S(0) $\lambda$28.2$\mu$m} \\
    \colhead{}     & \colhead{J2000} & \colhead{J2000} & \colhead{arcmin}     & 
    \colhead{$\times10^{-17}$ (W\,m$^{-2})$} &
    \colhead{$\times10^{-17}$ (W\,m$^{-2})$} &
    \colhead{$\times10^{-17}$ (W\,m$^{-2})$} 
  }
  \startdata
26  &  2:22:44.10 & +42:26:2.450  &   5.65  & $<$1.40        & $<$1.24       & $<$1.44        \\
25  &  2:22:43.35 & +42:25:33.85  &   5.53  & $<$1.47        & $<$1.39       & $<$2.10        \\
24  &  2:22:42.37 & +42:25:06.67  &   4.97  & $<$1.74        & $<$1.37       & 0.77$\pm$0.26   \\    		
23  &  2:22:41.44 & +42:24:40.09  &   4.42  & $<$1.67        & $<$1.58       & 2.41$\pm$0.63  \\	
22  &  2:22:40.78 & +42:24:07.65  &   3.87  & 0.67$\pm$0.11  & 1.45$\pm$0.11 & 5.08$\pm$0.62  \\	
21  &  2:22:39.83 & +42:23:39.90  &   3.32  & $<$1.89        & 1.77$\pm$0.30 & 5.77$\pm$0.85  \\	
20  &  2:22:38.38 & +42:23:13.93  &   2.76  & 1.09$\pm$0.09  & 2.49$\pm$0.16 & 8.84$\pm$1.07  \\	
19  &  2:22:37.59 & +42:22:46.22  &   2.21  & 4.11$\pm$0.45  & 4.10$\pm$0.48 & 10.91$\pm$0.68 \\
18  &  2:22:36.47 & +42:22:18.77  &   1.66  & 2.15$\pm$0.20  & 4.13$\pm$0.68 & 11.37$\pm$0.45 \\
17  &  2:22:35.47 & +42:21:50.29  &   1.11  & 2.79$\pm$0.63  & 3.82$\pm$0.50 & 11.07$\pm$0.48 \\
16  &  2:22:34.67 & +42:21:21.43  &   0.55  & 2.14$\pm$0.59  & 3.88$\pm$0.57 & 10.44$\pm$0.80 \\
1   &  2:22:33.41 & +42:20:56.88  &   0.00  & 3.43$\pm$0.19  & 6.86$\pm$0.57 & 14.64$\pm$0.58 \\
2   &  2:22:32.37 & +42:20:25.85  &  -0.55  & $<$1.58        & 4.35$\pm$0.59 & 11.66$\pm$0.54 \\	
3   &  2:22:31.02 & +42:19:59.96  &  -1.11  & 3.23$\pm$0.61  & 5.15$\pm$0.30 & 13.65$\pm$0.51 \\
4   &  2:22:30.16 & +42:19:31.63  &  -1.66  & 1.90$\pm$0.50  & 3.80$\pm$0.40 & 9.29$\pm$0.59  \\
5   &  2:22:29.15 & +42:19:03.34  &  -2.21  & 1.82$\pm$0.55  & 2.51$\pm$0.75 & 7.39$\pm$0.44  \\		
6   &  2:22:28.07 & +42:18:35.81  &  -2.76  & 1.28$\pm$0.11  & 2.81$\pm$0.21 & 6.82$\pm$0.68  \\	
7   &  2:22:26.99 & +42:18:08.74  &  -3.32  & $<$2.70        & 2.61$\pm$0.44 & 7.20$\pm$0.34  \\
8   &  2:22:25.98 & +42:17:41.93  &  -3.87  & $<$2.19        & 0.92$\pm$0.24 & 3.45$\pm$0.64  \\	
9   &  2:22:24.88 & +42:17:14.69  &  -4.42  & $<$2.40        & 1.58$\pm$0.50 & 4.24$\pm$0.53  \\	
10  &  2:22:23.74 & +42:16:46.54  &  -4.97  & $<$1.41        & 0.56$\pm$0.08 & 1.86$\pm$0.21  \\	
11  &  2:22:22.70 & +42:16:19.70  &  -5.53  & $<$1.64        & $<$1.05       & $<$2.24        \\	    	
12  &  2:22:21.65 & +42:15:52.55  &  -6.08  & $<$1.44        & $<$1.07       & $<$1.66        \\   		
\enddata
\tablenotetext{a}{Offset of each pointing along the disk of NGC\,891, from the pointing on the nucleus (position ID=1)}
\end{deluxetable}
              	            				
\newpage

\begin{deluxetable}{lccc}
  \tabletypesize{\scriptsize}
  \setlength{\tabcolsep}{0.02in}
  \tablecaption{Integrated Line Luminosities in NGC\,891\label{tab2}}
  \tablewidth{0pc}
  \tablehead{
    \colhead{Line}   & \colhead{Luminosity}    & \colhead{Corrected Luminosity\tablenotemark{a}}  & \colhead{Error} \\ 
    \colhead{} & \colhead{(L$_{\odot}$)}  & \colhead{(L$_{\odot}$)}  & \colhead{}   
    }
  \startdata
H$_2$ S(0)  &  5.44$\times 10^6$ & 7.18$\times 10^6$ &  24\%    \\
H$_2$ S(1)  &  8.94$\times 10^6$ & 1.32$\times 10^7$ &  24\%    \\
H$_2$ S(2)  &  5.26$\times 10^6$ & 7.80$\times 10^7$ &  32\%    \\	
\protect{[OI]} 63 $\mu$m  &  4.67$\times 10^7$ & 9.34$\times 10^7$ &  50\%    \\
\protect{[OI]} 146 $\mu$m  &  4.74$\times 10^6$ & 4.74$\times 10^6$ &  50\%    \\
\protect{[CII]} 158 $\mu$m  &  1.40$\times 10^8$ & 1.02$\times 10^7$ &  37\%    \\
L$_{FIR}$     &  1.51$\times 10^{10}$ & 1.13$\times 10^{10}$ &  30\%    \\
\enddata
\tablenotetext{a}{H$_2$ correction factor from extinction (1.16, 1.32, and 1.32 for the S(0), S(1) and S(2) lines respectively), and high latitude emission missed by our beam (1.16 correction factor). The [OI] 63 $\mu$m corrected by factor of 2 for geometry and [CII] line is corrected by 0.7 for HII region contribution. The FIR (42.5 to 122.5 $\mu$m) luminosity is divided by two to account for cirrus emission, then multiplied by 1.5 for flux longward of 122.5 $\mu$m as required by the PDRT models. The factor of 1.5 is appropriate for galaxies in the Revised Bright Galaxy Survey \citep{Sanders03}}
\end{deluxetable}
        
\newpage

\begin{figure}
\epsscale{1.0}
 \plotone{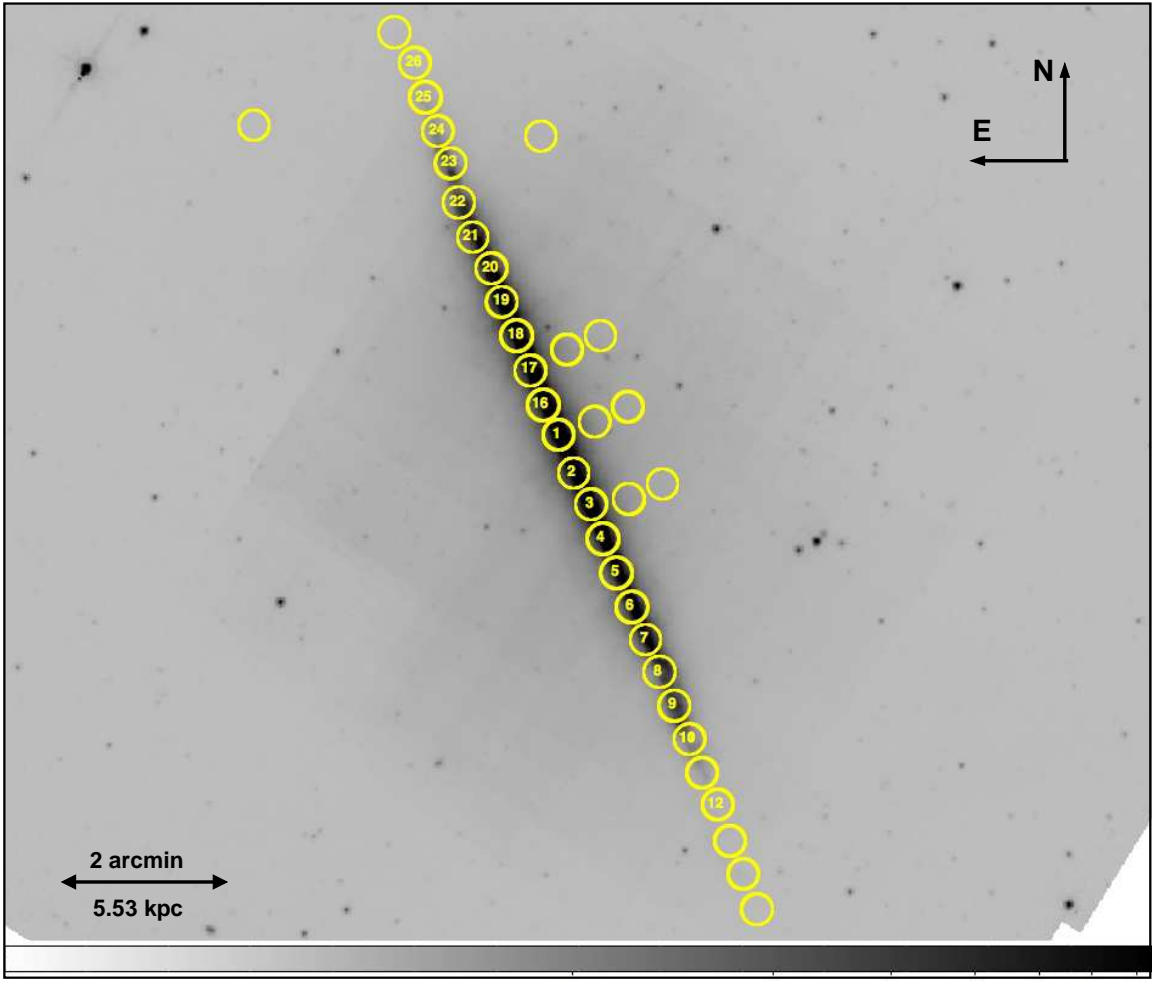}
  \caption{A Spitzer/IRAC 8$\mu$m image of the NGC\,891. The IRS pointings are indicated with circles of 25$''$ in  diameter. The positions where at least one H$_2$ line was detected and listed in Table 1 are marked with their corresponding number.}
  \label{fig:fig1}
\end{figure}

\begin{figure}
\epsscale{0.78}
  \plotone{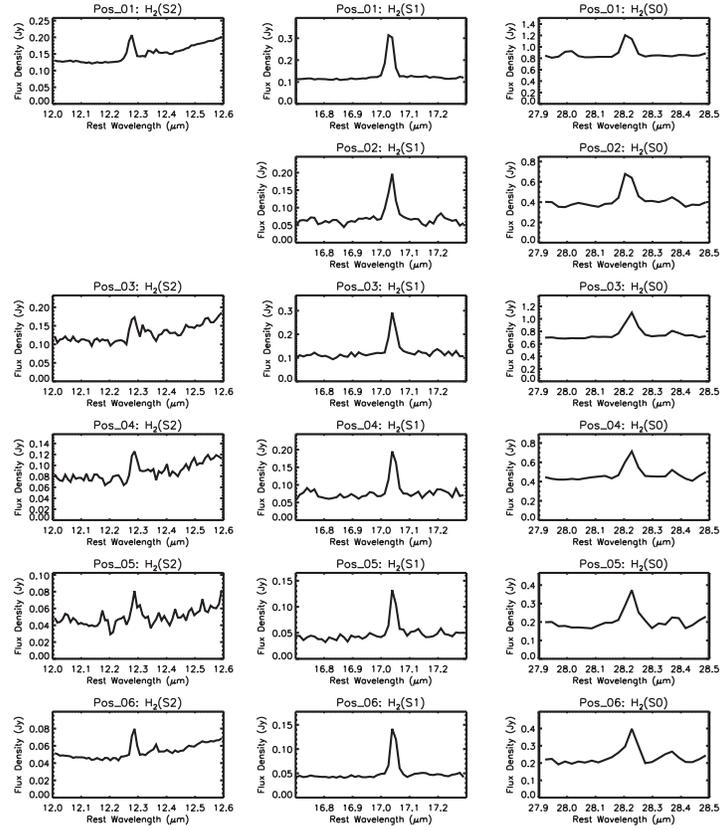}
  \caption{Zoomed regions near the H$_2$ lines of the final IRS spectra at each one at the pointings listed in Table 1, where at least one of the S(0), S(1), and S(3) lines were detected. The spectra are arranged so the left, center, and right columns refer to the S(2), S(1) and S(0) lines respectively, and each row refers to a position on the sky (as listed in Table 1)}.
\end{figure}

\begin{figure}
  \setcounter{figure}{1}
\epsscale{0.78}
  \plotone{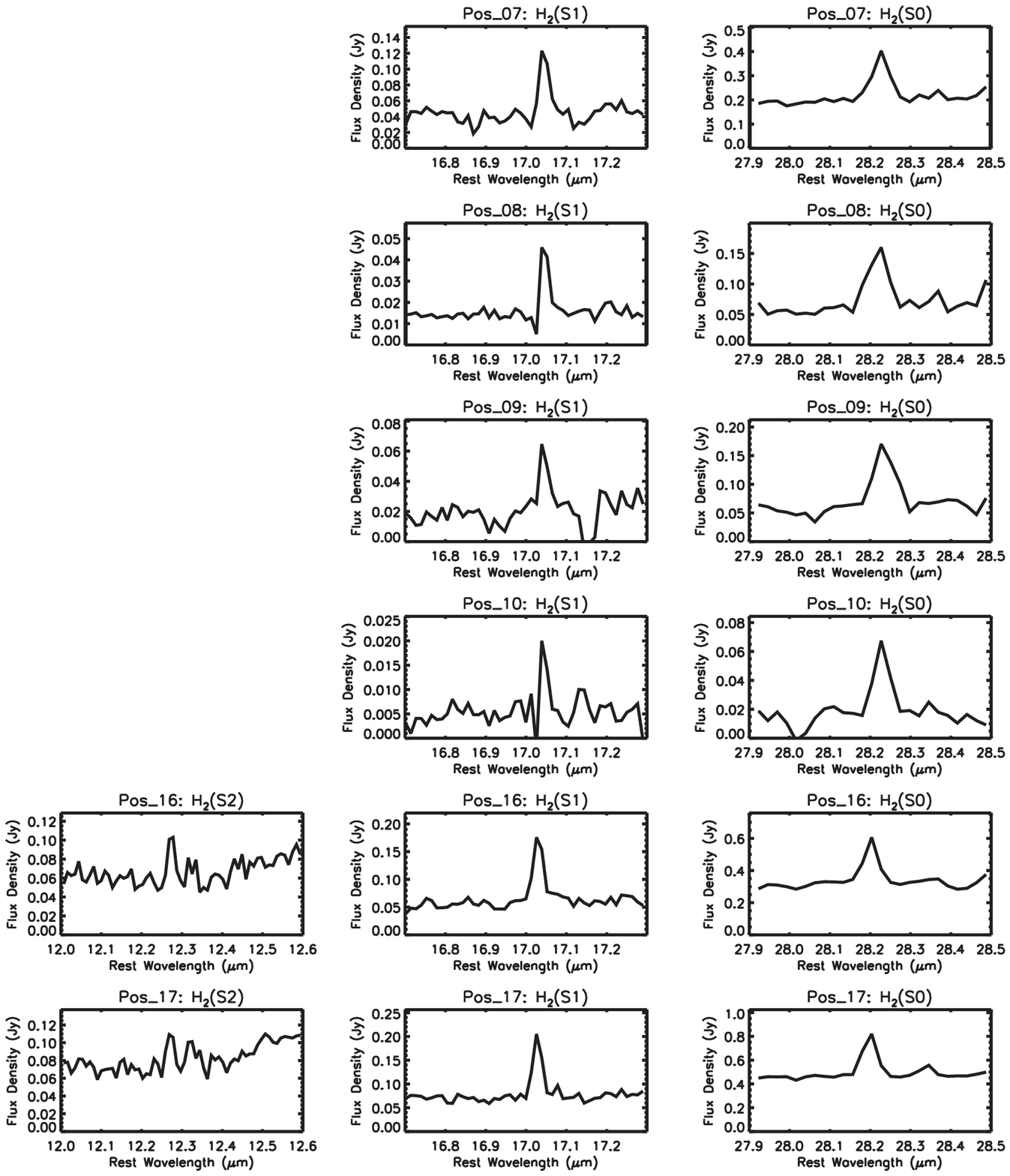}
  \caption{Continued.}
\end{figure}

\begin{figure}
 \setcounter{figure}{1}
\epsscale{0.78}
  \plotone{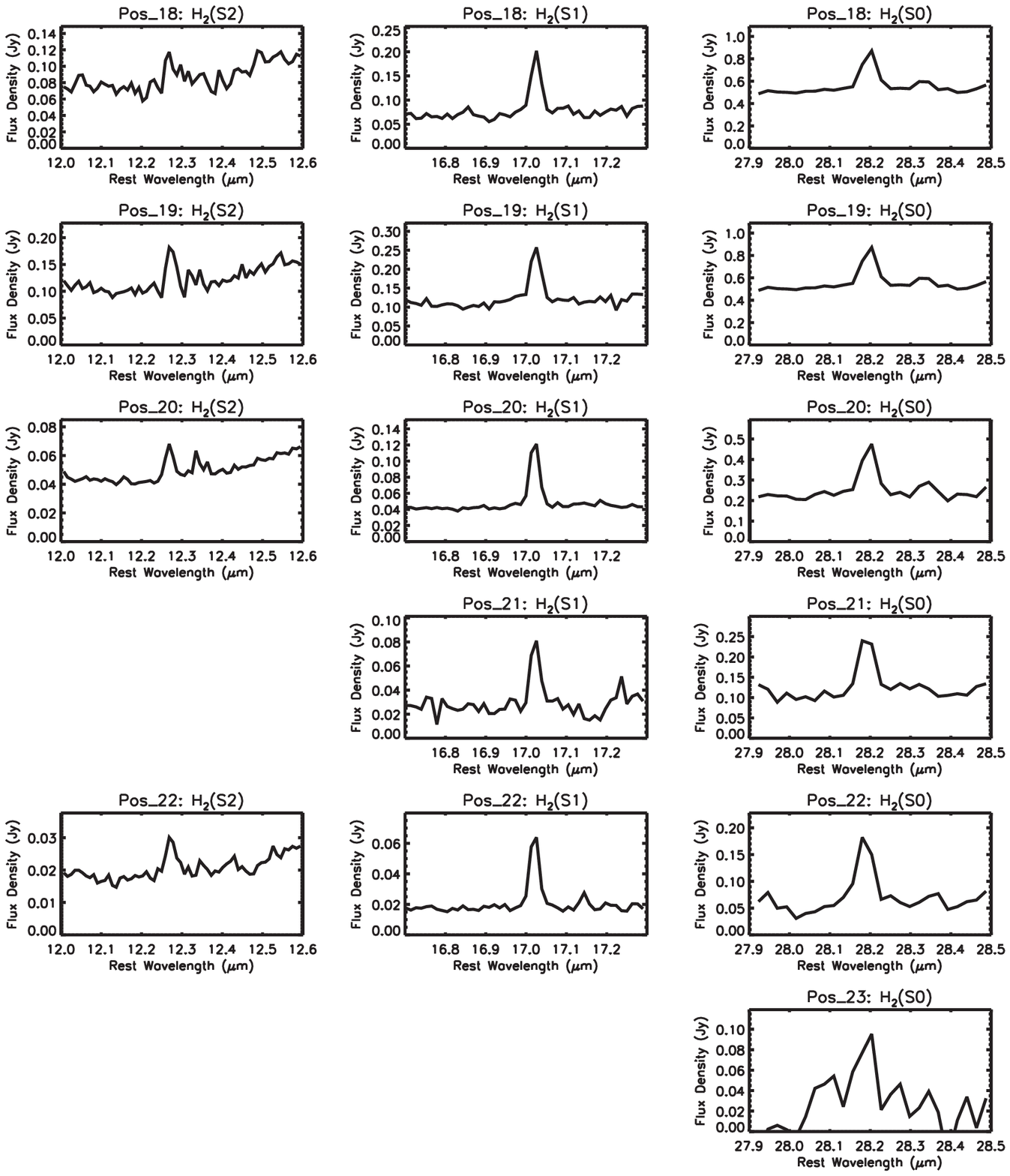}
  \caption{Continued.}
\end{figure}

\begin{figure}
 \setcounter{figure}{1}
\epsscale{0.78}
  \plotone{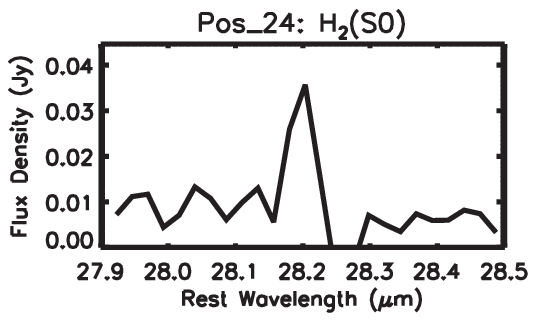}
  \caption{Continued.}
\end{figure}

\begin{figure}
\epsscale{0.9}
\plotone{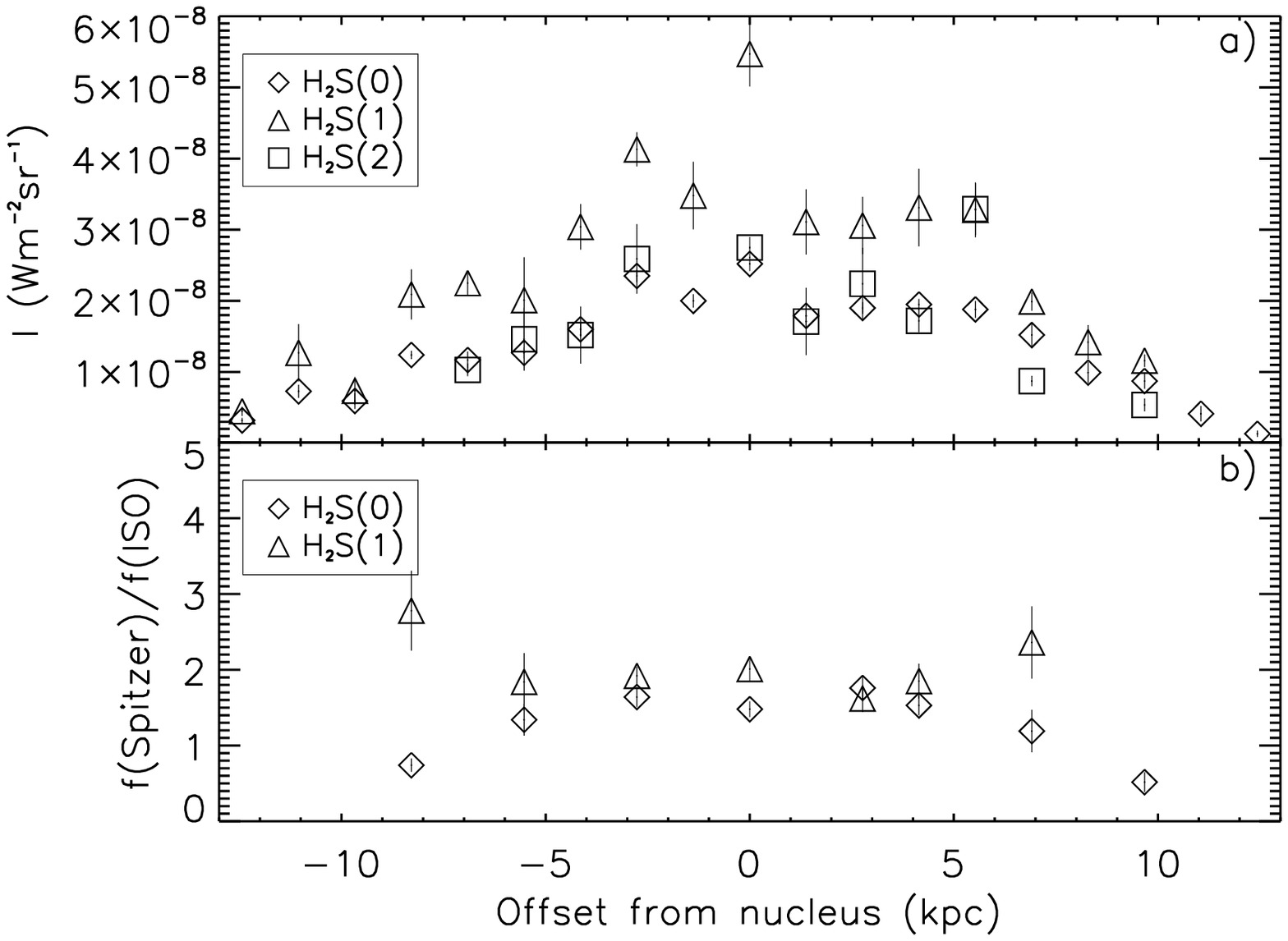}
  \caption{a) Observed Spitzer H$_{2}$ line intensities as a function of distance from the nucleus. Positive offsets are to the north-east.  b) Observed Spitzer/ISO S(0), and S(1) line flux ratios normalized to the ISO beam sizes.}
\end{figure}

\begin{figure}
\epsscale{0.9}
\plotone{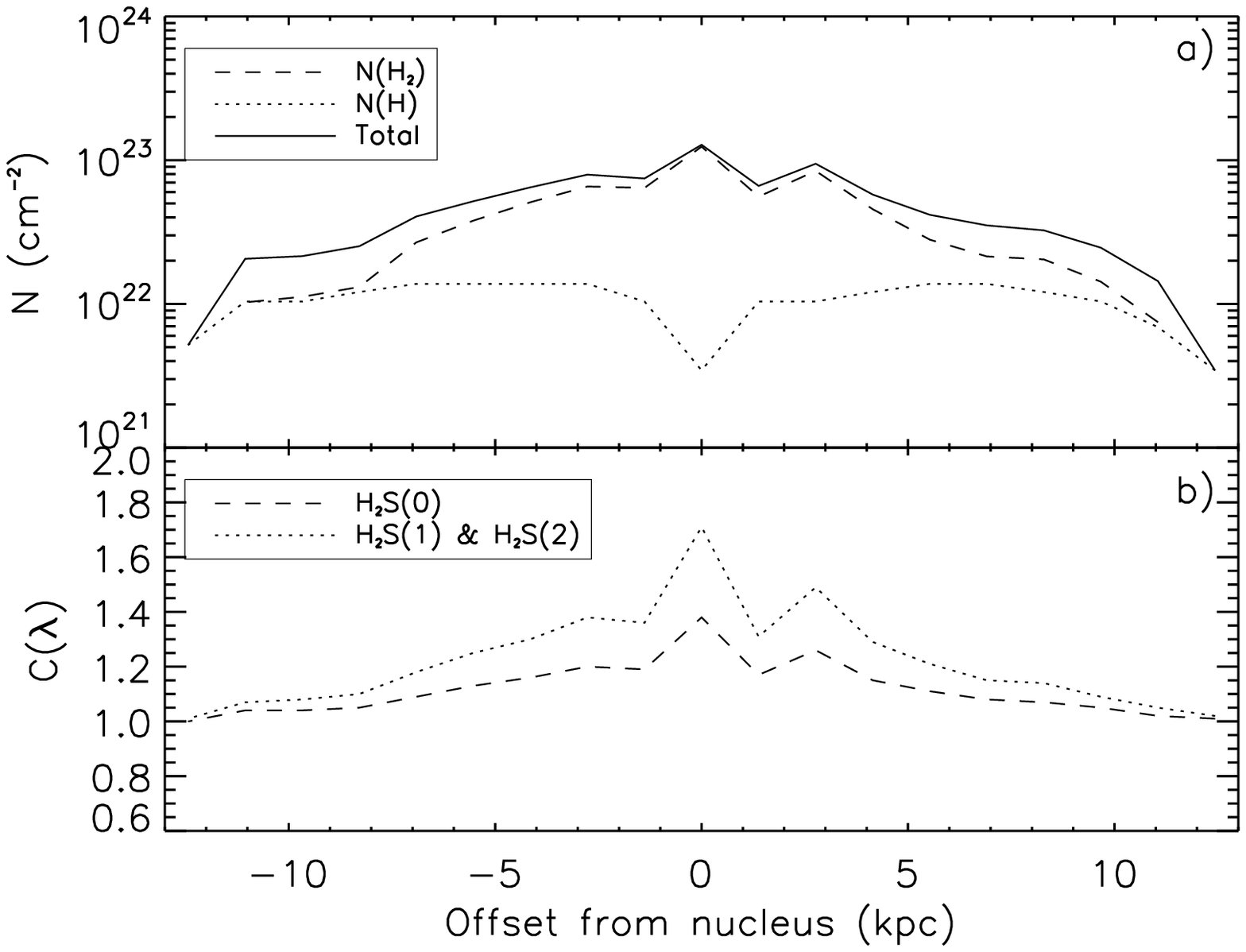}
   \caption{a) The column densities of hydrogen nuclei contained in H$_2$ (derived from CO(1-0), see section 3.3), HI, and the sum of the two as a function of offset (in kpc) along the galactic plane from the nucleus. b): Extinction correction, C($\lambda$), for the observed molecular hydrogen line emission. Note that the correction is the same for the H$_2$S(1) and H$_2$S(2) lines.}
\end{figure}

\begin{figure}
\epsscale{0.9}
\plotone{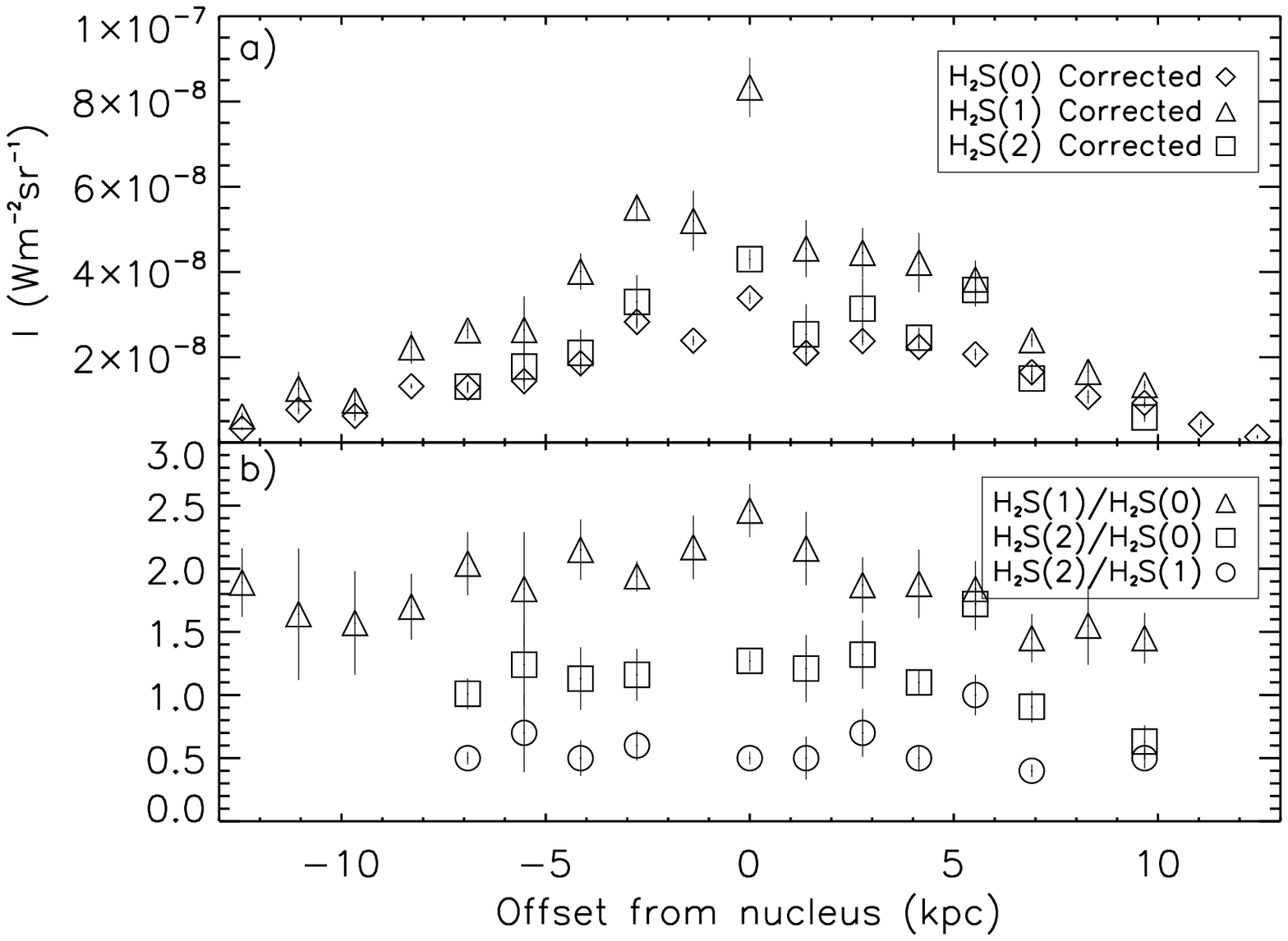}
  \caption{a) Extinction corrected Spitzer H$_{2}$ line intensities as a function of distance from the nucleus. b) S(1)/S(0), S(2)/S(0) and S(2)/S(1) line intensity ratios calculated from the extinction corrected data in panel a.}
\end{figure}

\begin{figure}
\epsscale{0.9}
\plotone{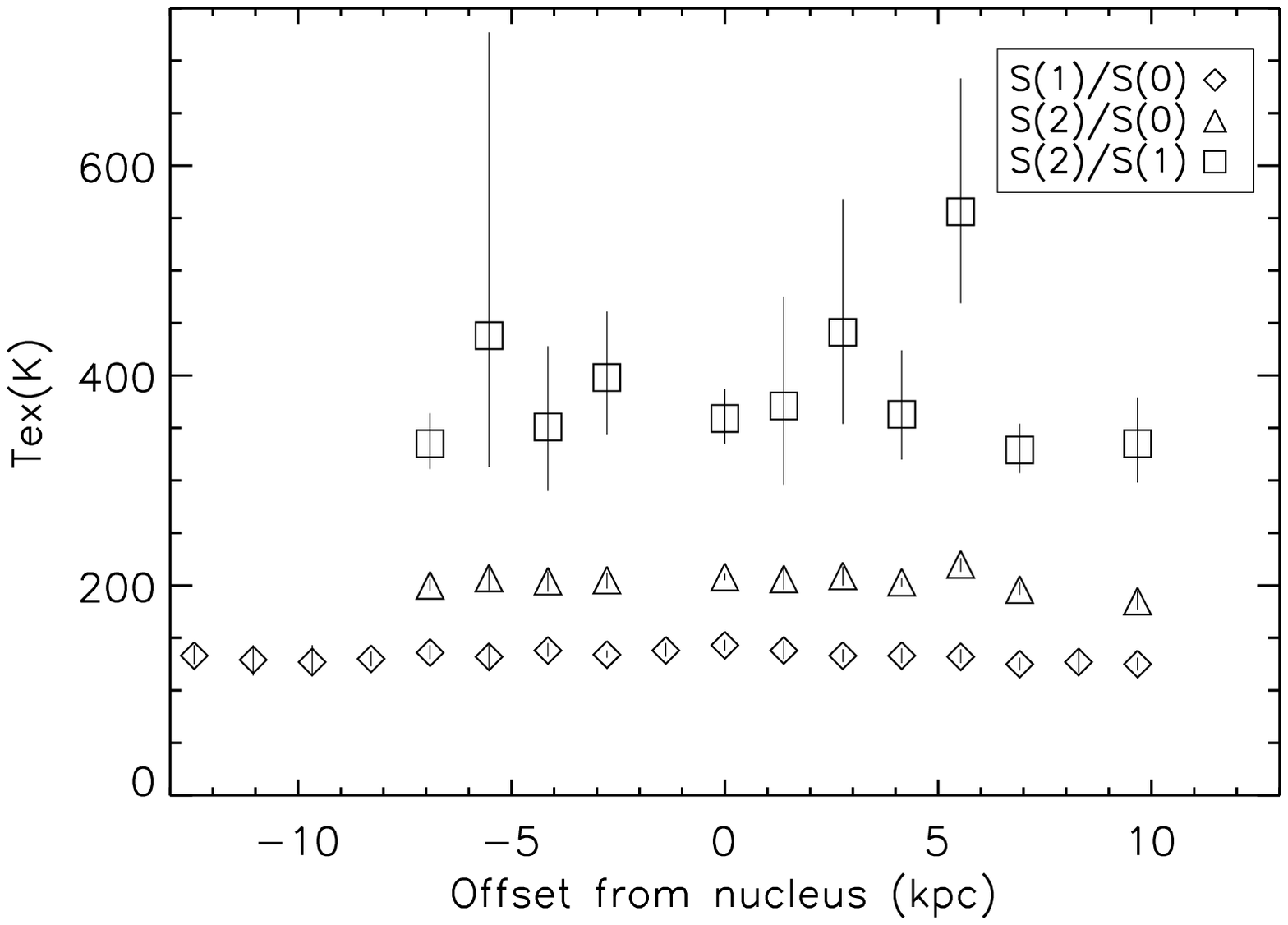}
  \caption{Level excitation temperatures determined from the extinction corrected line intensity ratios.  For T$_{ex}$(1,0) we assume the $o/p$ ratio is given by thermodynamic equilibrium at T = 135 K: $o/p$ = 2.3.}
\end{figure}

\begin{figure}
\epsscale{0.9}
\plotone{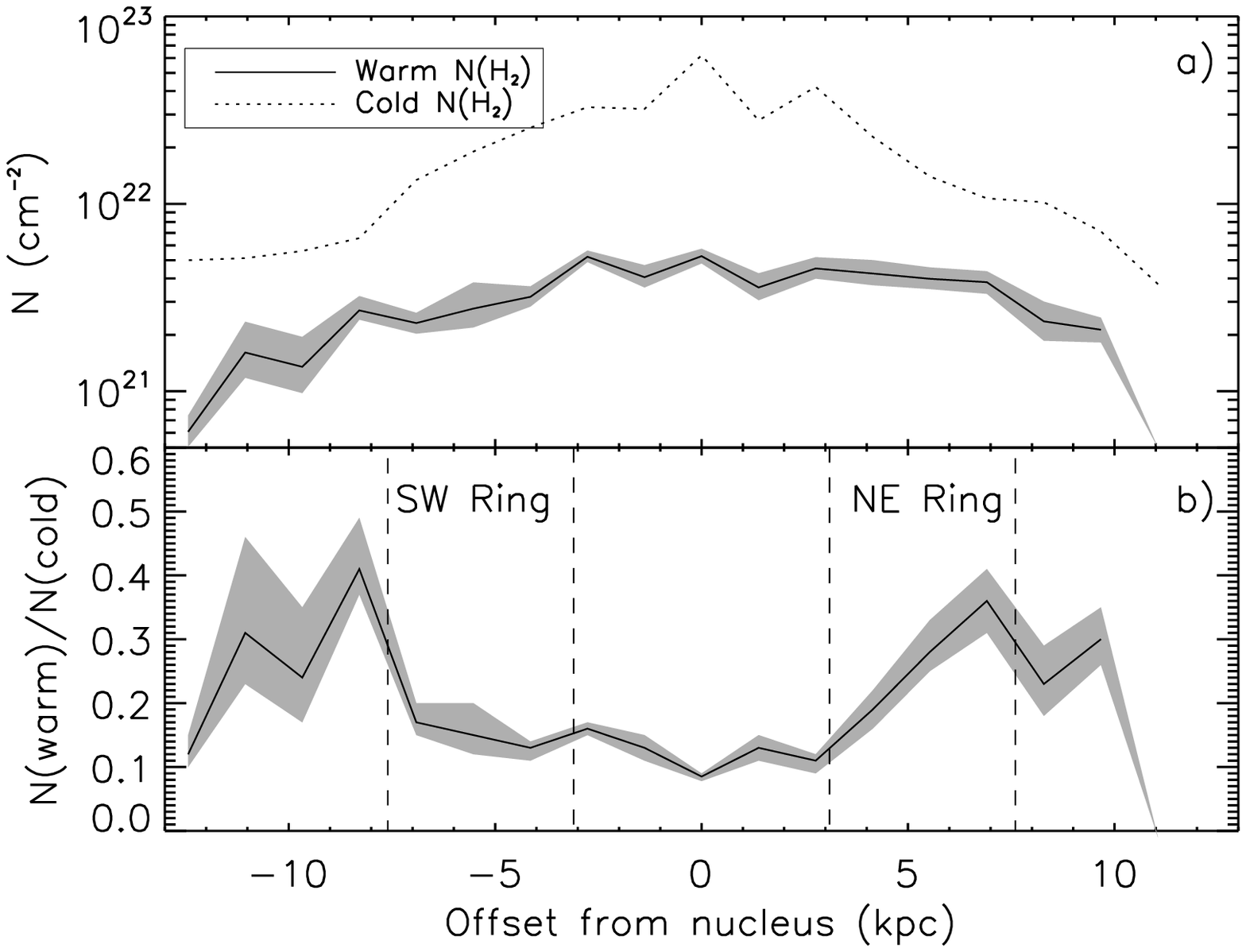}
  \caption{a) Column densities in H$_2$ as traced by the CO(1-0) line (dotted line, ``cold'' H$_2$) and as traced by the S(0) and S(1) lines (solid line, ``warm'' H$_2$).  Error bounds are in grey.  b) The ratio of ``warm'' to ``cold'' H$_2$ column densities.  We have enclosed the regions of the NE and SE ring by dotted lines.}
\end{figure}

\begin{figure}
\epsscale{0.9}
\plotone{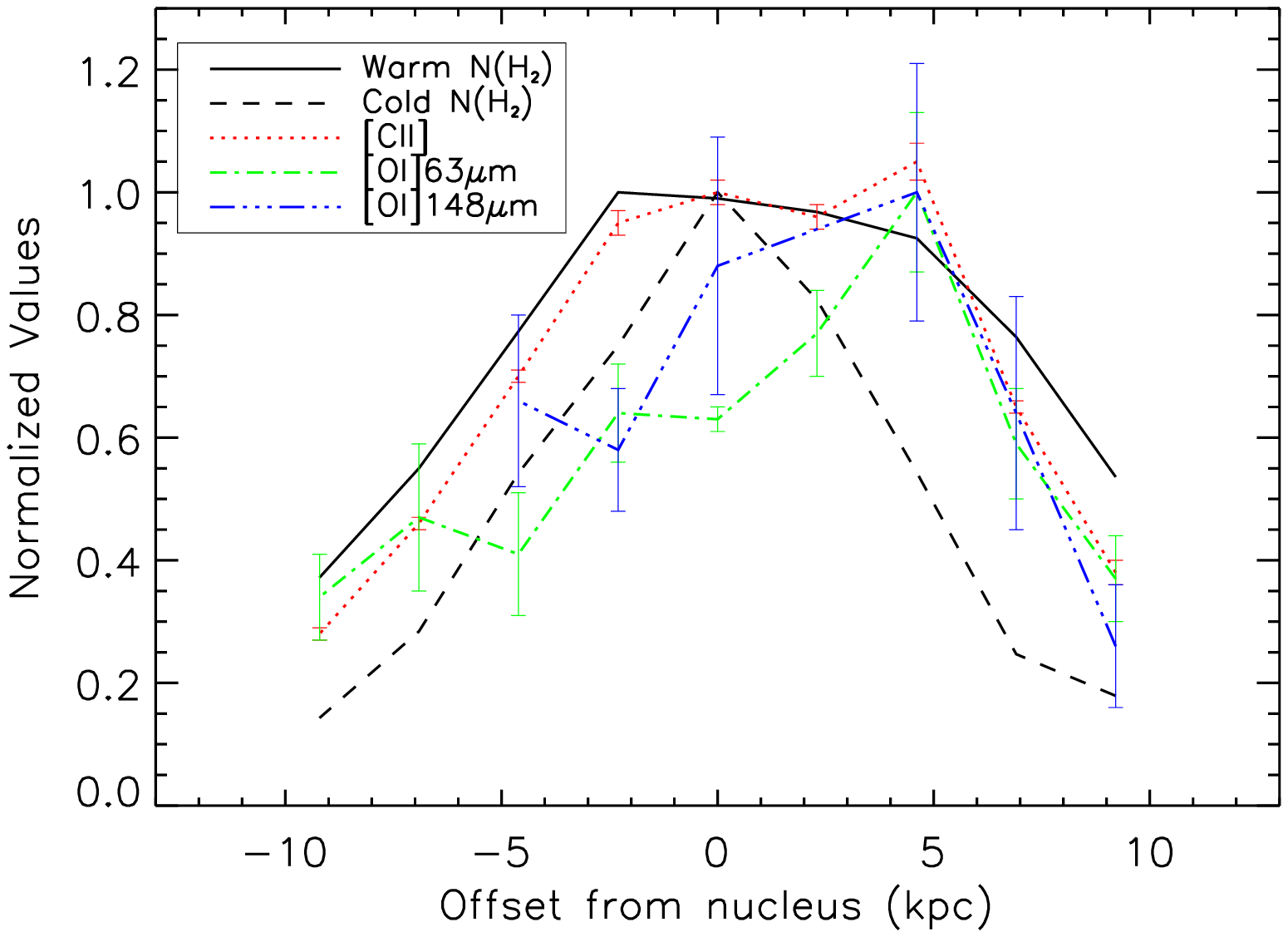}
  \caption{Cuts in the [CII] (158 $\mu$m), and [OI] (63 and 146 $\mu$m) lines along the plane of NGC\,891 (from the ISO satellite, \citet{Brauher08}), superposed on our derived cuts in the ``warm" and ``cold" H$_2$ gas which have been smoothed to the spatial resolution of the ISO beam (75'').  There is excellent correspondence between the warm H$_2$ and the [CII] line including the asymmetry to the north of the nucleus strongly suggesting a PDR origin for the H$_2$ line emission.}
\end{figure}

\begin{figure}
\epsscale{1}
\plotone{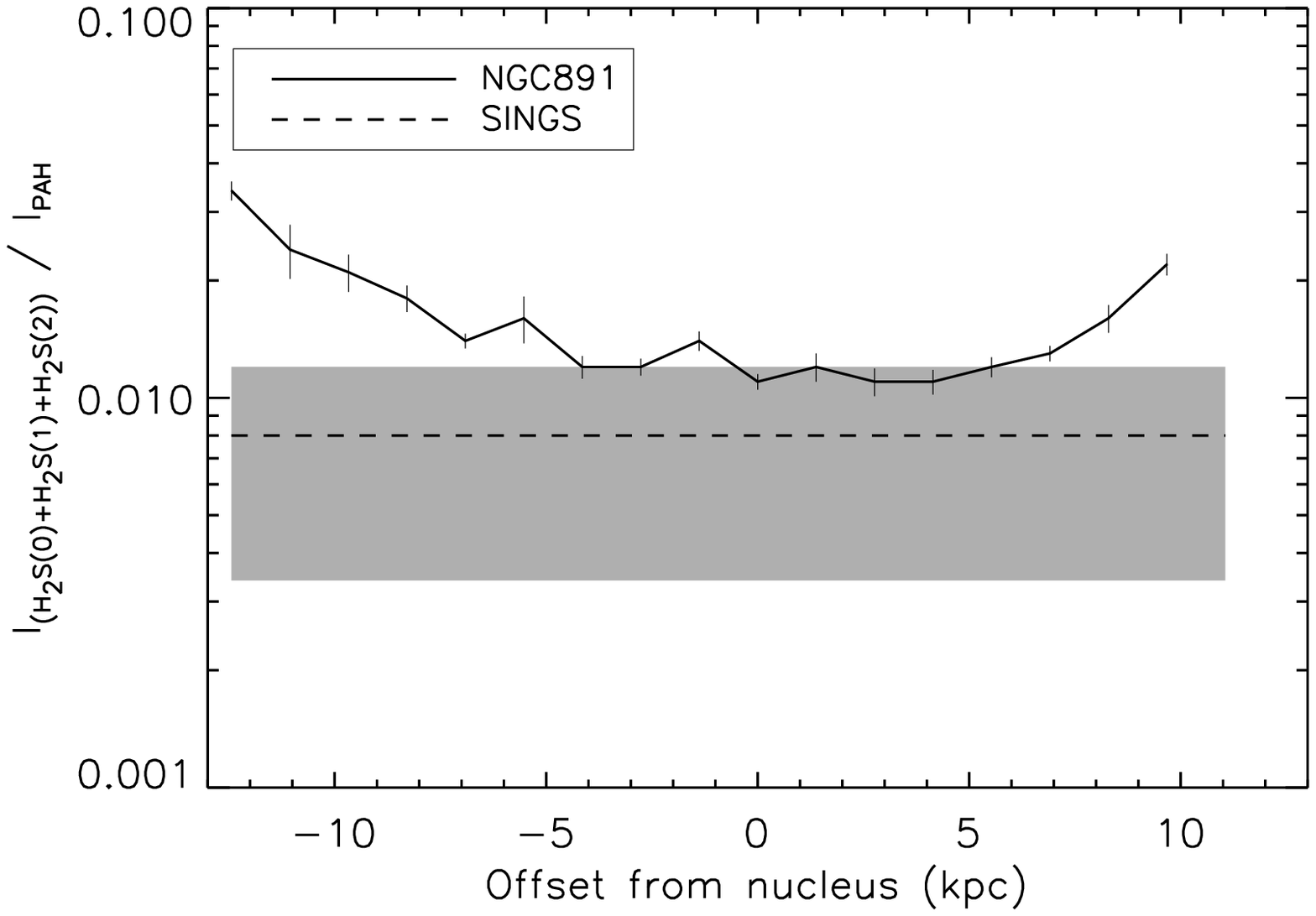}
  \caption{Ratio between the H$_2$ line emission, summed over the S(0), S(1) and S(2) lines and the PAH $7.7\mu$m feature emission (measured from the IRAC 8$\mu$m imaging data as in \citet{Roussel07}) along the plane of NGC\,891. The ratio is remarkably constant though the inner parts of the galaxy and rises by about a factor 2 at the profile edges. The range of values measured in SINGS star forming galaxies (mean value and 1$\sigma$ scatter) is shown for comparison.  }
\end{figure}

\begin{figure}
\epsscale{1}
  \plotone{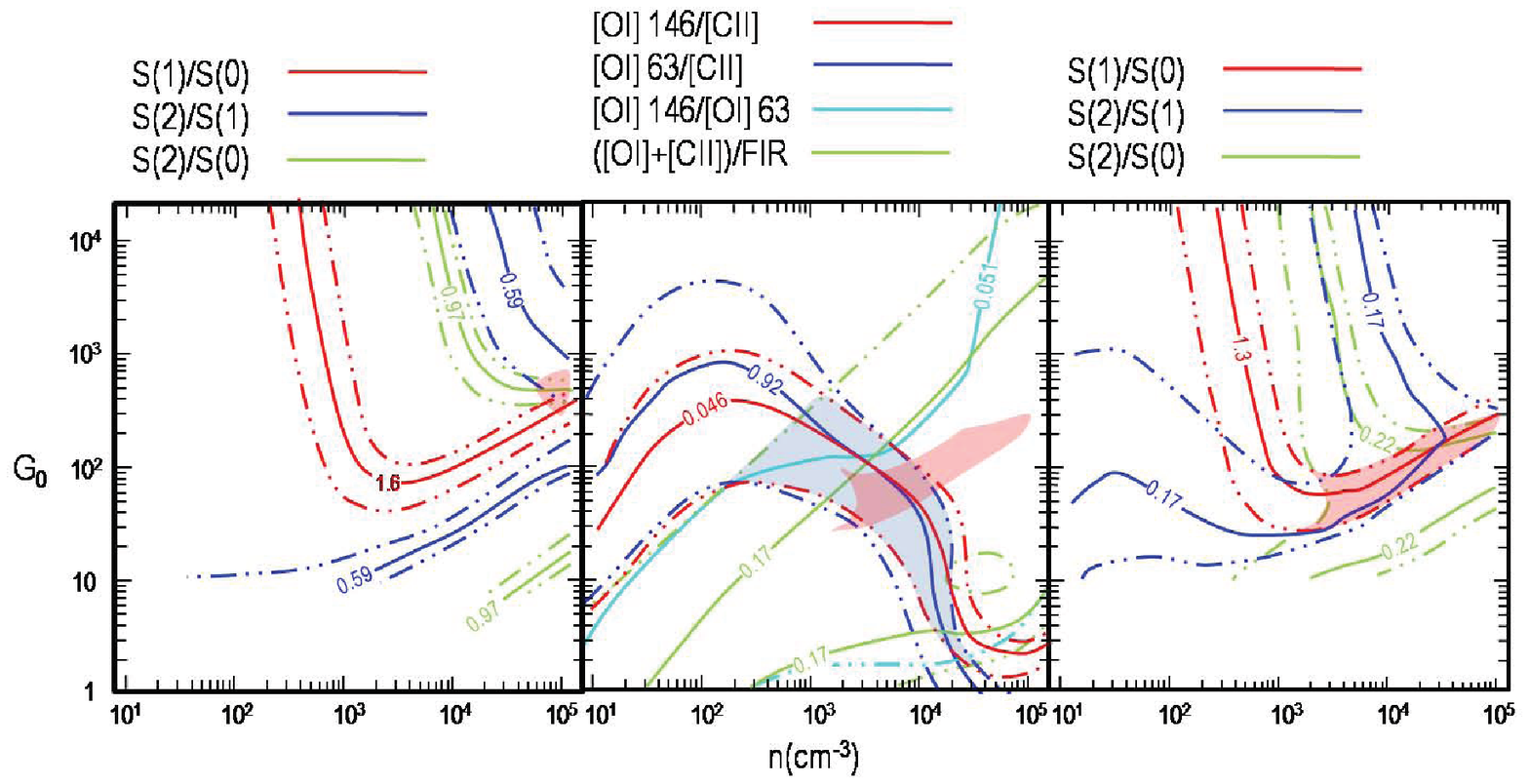}
  \caption{ PDR solutions for the G and n from the Photodissciation Region Toolshead (PDRT) (http://dustem.astro.umd.edu/).  (left)  Solutions for the H$_2$ lines.  The red shaded region indicates allowed solutions when all of the line emission is from PDRs.  (center) Solutions for the FIR lines and the ([OI]+[CII])/FIR continuum luminosity ratios.  The blue shaded area indicates the allowed solutions with the various corrects to the tracers described in the text. The red shaded area is the solution for the H$_2$ lines from the PDR/shock model on the right. (right) Solutions for the H$_2$ lines after subtracting off a shock contribution.  The red shaded region indicates allowed solutions when 80\% of the S(2) line emission, and 30\% of the S(1) line emission is assumed to arise from shocks.}
\end{figure}

\begin{enumerate}
	\item 
\end{enumerate}

\end{document}